\documentclass[aps,prd,amsmath,amssymb,amsfonts,onecolumn,nofootinbib,preprint]{revtex4}
\usepackage{graphicx}% Include figure files
\usepackage{dcolumn}% Align table columns on decimal point
\usepackage{bm}% bold math
\usepackage{amsmath}
\usepackage{amssymb}
\usepackage{color}

\begin{document}

\title{Quantum corrected phase diagram of holographic fermions% :\\
 % AdS electron stars with Dirac hair
}

\author{Mariya~V.~Medvedyeva}
\affiliation{Department of Physics, Georg-August-Universit\"{a}t G\"{o}ttingen,\\
Friedrich-Hund-Platz 1, 37077 G\"{o}ttingen, Germany}
\author{Elena Gubankova}
\affiliation{Institute for Theoretical Physics, J.~W.~Goethe-University,\\
D-60438 Frankfurt am Main, Germany}
\author{Mihailo \v{C}ubrovi\'c}
\affiliation{Instituut Lorentz, Delta-Institute for Theoretical Physics, Leiden University, Niels Bohrweg 2,\\
2300 RA Leiden, Netherlands}
\author{Koenraad Schalm}
\affiliation{Instituut Lorentz, Delta-Institute for Theoretical Physics, Leiden University, Niels Bohrweg 2,\\
2300 RA Leiden, Netherlands}
\author{Jan Zaanen}
\affiliation{Instituut Lorentz, Delta-Institute for Theoretical Physics, Leiden University, Niels Bohrweg 2,\\
2300 RA Leiden, Netherlands}

\begin{abstract}
We study the phases of strongly correlated electron systems in two spatial dimensions in the framework of AdS${}_4$/CFT${}_3$ correspondence. The AdS (gravity) model consists of a Dirac fermion coupled to electromagnetic field and gravity. To classify the ground states of strongly correlated electrons on the CFT side and to construct the full phase diagram of the system, we construct a quantum many-body model of bulk fermion dynamics, based on the WKB approximation to the Dirac equation. At low temperatures, we find a quantum corrected approximation to the electron star where the edge is resolved in terms of wave functions extended fully through AdS. At high temperatures, the system exhibits a {\em first} order thermal phase transition to a charged AdS-RN black hole in the bulk and the emergence of local quantum criticality on the CFT side. This change from the third order transition experienced by the semi-classical electron star restores the intuition that the transition between the critical AdS-RN liquid and the finite density Fermi system is of van der Waals liquid-gas type.
\end{abstract}

\maketitle

\section{Introduction}

The problem of fermionic quantum criticality has proven hard enough for condensed matter physics to keep seeking new angles of attack. The main problem we face is that the energy scales vary by orders of magnitude between different phases. The macroscopic, measurable quantities emerge as a result of complex collective phenomena and are difficult to relate to the microscopic parameters of the system. An illustrative example present the heavy fermion materials \cite{heavy} which still behave as Fermi liquids but with vastly (sometimes hundredfold) renormalized effective masses. On the other hand, the strange metal phase of cuprate-based superconducting materials \cite{strange}, while remarkably stable over a range of doping concentrations, shows distinctly non-Fermi liquid behavior. The condensed matter problems listed all converge toward a single main question in field-theoretical language. It is the classification of ground states of interacting fermions at finite density.

In this paper we attempt to understand these ground states in the framework of AdS/CFT, the duality between the strongly coupled field theories in $d$ dimensions and a string configuration in $d+1$ dimension. Holography (AdS/CFT correspondence) \cite{Hartnoll:2010,McGreevy:2010} has become a well-established treatment of strongly correlated electrons by now, but it still has its perplexities and shortcomings. Since the existence of holographic duals to Fermi surfaces has been shown in \cite{Vegh:2009,Leiden:2009}, the next logical step is to achieve the understanding of the phase diagram: what are the stable phases of matter as predicted by holography, how do they transform into each other and, ultimately, can we make predictions on quantum critical behavior of real-world materials based on AdS/CFT.

The classification of ground states now translates into the following question: classify the stable asymptotically AdS geometries with charged fermionic matter in a black hole background. Most of the work done so far on AdS/CFT for strongly interacting fermions relies on bottom-up toy gravity models and does not employ a top-down string action. We stay with the same reasoning and so will work with Einstein gravity in $3+1$ dimensions. We note, however, that top-down constructions of holographic fermions exist \cite{Gubser,GauntlettSonner}.

In this paper we construct a model dubbed "WKB star", alluding to the fact that we treat the same large occupation number limit as the electron star \cite{Hartnoll:es} but go further from the ideal fluid limit of \cite{Hartnoll:es}. The main idea is to solve the fermionic equations of motion in the WKB limit \emph{without} taking the fluid limit: the total density is the sum of the contributions of individual wave functions rather than an integral over them. The main approximation we introduce is thus just the quasiclassical treatment of fermions, inherent to WKB. The inverse occupation number serves as the control parameter of this approximation. In addition, we assume that the correction to the fluid limit is captured by the correction to the pressure. This assumption cannot be rigorously derived. We will discuss, however, the robustness of our findings with respect to this assumption. In addition to simply improving the mathematical treatment of the bulk many-body fermion system, we will show that some properties of the system change nonperturbatively in the fluid limit. In particular, the thermodynamic behavior of the system at finite temperature is changed compared to the electron star.

We will use a simple WKB formalism to approximate the many-body Fermi system in the AdS bulk. This adds quantum corrections to the Thomas-Fermi (fluid) approximation by taking into account finite level spacing. In other words, we do not take the limit of an infinite number of occupied levels but keep the occupation number finite. The occupation number itself acts as the control parameter of our approximation. The most notable feature, however, occurs in the transition from the semiclassical approximation at {\em infinite} occupation number to finite occupation number. We find that the finite density quantum many body phases with fermionic quasiparticles at high enough temperatures always exhibit a {\em first} order transition into the zero density AdS-RN phase. Intuitively, this can be interpreted as a universal van der Waals liquid-gas transition. On the other hand in the semiclassical fluid limit underlying the electron star, the transition was found to be continuous \cite{Hartnoll:phtr,Larus}. With this re-emergence of the first order nature of the thermal phase transition at the quantum level our results confirm the intuition that a density driven phase transition is always first order as also indicated by the Dirac hair approximation \cite{csz2010}. We thus show with an explicit calculation that in the context of fermionic questions in AdS/CFT quantum ``1/N'' corrections can be important and that the semiclassical fluid limit can be unreliable, at least at finite temperature. While the quantum corrections likely have important consequences also at $T=0$, we have not explored the zero-temperature physics in this paper.

The outline of the paper is as follows. In the Section II we describe the field content and geometry of our gravity setup, an Einstein-Maxwell-Dirac system in $3+1$ dimension, and review the single-particle solution to the bulk Dirac equation. In Section III we start from that solution and apply the WKB approximation to derive the Dirac wave function of a many-particle state in the bulk. Afterwards we calculate density and pressure of the bulk fermions -- the semiclassical estimate and the quantum corrections, thus arriving at the equation of state. Section IV contains the numerically self-consistent solution of the set of equations for fermions, gauge field and the metric. There we also describe our numerical procedure. Section V is the core, where we analyze thermodynamics and spectra of the field theory side and identify different phases as a function of the three parameters of the system: chemical potential $\mu$, fermion charge $e$ and conformal dimension $\Delta$. Section VI sums up the conclusions and offers some insight into possible broader consequences of our work and into future steps.

\section{Holographic fermions in charged background}

We wish to construct the gravity dual to a field theory at finite fermion density. We will specialize to $2+1$-dimensional conformal systems of electron matter, dual to AdS${}_4$ gravities. We consider a Dirac fermion of charge $e$ and mass $m$ in an electrically charged gravitational background with asymptotic AdS geometry. Adopting the AdS radius as the unit length, we can rescale the metric $g_{\mu\nu}$ and the gauge field $A_\mu$:
\begin{equation}
\label{rescale}g_{\mu\nu}\mapsto g_{\mu\nu}L^2,~~ A_\mu\mapsto LA_\mu.
\end{equation}
In these units, the action of the system is:
\begin{equation}
\label{action}S=\int d^4x\sqrt{-g}\left[\frac{1}{2\kappa^2}L^2\left(R+6\right)+\frac{L^2}{4}F^2+L^3\mathcal{L}_f\right]
\end{equation}
where $\kappa$ is the gravitational coupling and $F_{\mu\nu}=\partial_\mu A_\nu-\partial_\nu A_\mu$ is the field strength tensor. The fermionic Lagrangian is:
\begin{equation}
\label{lagfermi}\mathcal{L}_f=\bar{\Psi}\left[e^\mu_A\Gamma^A\left(\partial_\mu+\frac{1}{4}\omega_\mu^{BC}\Gamma_{BC}-ieLA_\mu\right)-mL\right]\Psi
\end{equation}
where $\bar{\Psi}=i\Psi^\dagger\Gamma^0$, $e^\mu_A$ is the vierbein and $\omega_\mu^{AB}$ is the spin connection.

We shall be interested in asymptotically AdS solutions with an electric field. The $U(1)$ gauge field is simply $A=\Phi dt$ and we parametrize our  metric in four spacetime dimensions as:
\begin{equation}
\label{metric}ds^2=\frac{f(z)e^{-h(z)}}{z^2}dt^2-\frac{1}{z^2}\left(dx^2+dy^2\right)-\frac{1}{f(z)z^2}dz^2
\end{equation}
The radial coordinate is defined for $z\geq 0$, where $z=0$ is the location of AdS boundary. All coordinates are dimensionless, according to (\ref{rescale}).
This form of the metric is sufficiently general to model any configuration of static and isotropic charged matter. Development of a horizon at finite $z$ is signified by the appearance of a zero of the function $f(z)$, $f(z_H)=0$. From now on we will set $L=1$.

We will now proceed to derive the equation of motion for the Dirac field. From (\ref{lagfermi}), the equation reads:
\begin{equation}
\label{diraceq}e^\mu_A\Gamma^A\left(\partial_\mu+\frac{1}{4}\omega_\mu^{BC}\Gamma_{BC}-ieA_\mu\right)\Psi=m\Psi.
\end{equation}
In the metric (\ref{metric}) we can always eliminate the spin connection \cite{Vegh:2009} by transforming:
\begin{equation}
\label{chitrans}\Psi\mapsto
(gg^{zz})^{-\frac{1}{4}}\Psi=\frac{e^{h(z)/4}z^{3/2}}{f(z)^{1/4}}\Psi\equiv a^{-1}(z)\Psi.
\end{equation}
At this point it is convenient to adopt a specific representation of gamma matrices. We choose:
\begin{equation}
\Gamma^0=\left(\begin{matrix}1 & 0\\ 0 & -1\end{matrix}\right),~~\Gamma^{x,y,z}=\left(\begin{matrix}0 & \sigma_{1,2,3}\\ -\sigma_{1,2,3} & 0\end{matrix}\right).
\end{equation}
In this basis we define the radial projections $\Psi_\pm$ as eigenvalues of the projection operator onto the time axis:
\begin{equation}
\label{proj}\Psi_\pm=\frac{1}{2}\left(1\pm\Gamma^0\right)\Psi,
\end{equation}
after which the Dirac equation in matrix form becomes:
\begin{equation}
\label{direqmat}
\sqrt{f}\partial_z\left(\begin{matrix}\Psi_+\\
\Psi_-\end{matrix}\right)=\hat{D}\left(\begin{matrix}\Psi_+\\
\Psi_-\end{matrix}\right)~.
\end{equation}
Here the matrix $\hat{D}$ is the differential operator along the transverse coordinates ($x,y$) and time, which we will specify shortly.

We will now set the stage for solution of the Dirac equation in the WKB approximation. We can separate the radial dynamics (along the $z$ coordinate) from the motion in the $x-y$ plane. We can thus make the separation ansatz:
\begin{equation}
\label{cylansatz}\left(\begin{matrix}\Psi_+(t,z,x,y)\\ \Psi_-(t,z,x,y)\end{matrix}\right)=\int\frac{d\omega}{2\pi}\left(\begin{matrix}F(z)K_1(x,y)\\
-G(z)K_2(x,y)\end{matrix}\right)e^{-i\omega t}
\end{equation}
where the $F,G$ are scalars and the modes $K_{1,2}$ are in-plane spinors. The Dirac equation then takes the form:
\begin{equation}
\label{direqcyl}\left(\begin{matrix}\partial_zF K_1\\
-\partial_zG
K_2\end{matrix}\right)=\left(\begin{matrix}-\hat{\partial}/\sqrt{f(z)} & \left(\tilde{E}\left(\omega,z\right)+\tilde{M}\left(z\right)\right)\sigma_3\\ \left(\tilde{E}\left(\omega,z\right)-\tilde{M}\left(z\right)\right)\sigma_3 & -\hat{\partial}/\sqrt{f(z)}\end{matrix}\right)\left(\begin{matrix}FK_1\\
-GK_2\end{matrix}\right)
\end{equation}
We recognize the matrix at the right hand side as $\hat{D}/\sqrt{f}$. The terms $\tilde{E}$ and $\tilde{M}$ have the meaning of local energy and mass terms, respectively:
\begin{equation}
\label{mbdef}
\tilde{E}(z)=-\frac{e^{h\left(z\right)/2}}{f(z)}(\omega+e\Phi(z)),~\tilde{M}(z)=\frac{m}{z\sqrt{f(z)}}.
\end{equation}
The in-plane operator $\hat{\partial}$ acts on each in-plane spinor as:
\begin{equation}
\label{part}\hat{\partial}=\left(\begin{matrix}0 &
i\bar{\partial}\\ -i\partial & 0\end{matrix}\right)
\end{equation}
with $\partial\equiv\partial_x+i\partial_y$. To maintain the separation of variables in (\ref{direqcyl}), we require $\hat{\partial}K_i=\lambda_iK_i$, where $\vert\lambda_i\vert^2$ corresponds the momentum-squared of the in-plane motion of the particle. The physical requirement that this momentum be the same for both radial projections translates into the condition $|\lambda_2|=|\lambda_1|$. Consistency of the separation of variables then shows us that $K_2=\sigma_3K_1$ and thus $\lambda_1=-\lambda_2=k$. This solves the $x,y$-dependent part of the equation, in terms of $\rho\equiv\sqrt{x^2+y^2}$ and $\phi=\arctan y/x$:
\begin{equation}
\label{ksol}K_i(x,y)=\left(\begin{matrix}J_{l-1/2}(\lambda_i\rho)e^{i(l-1/2)\phi} \\ J_{l+1/2}(\lambda_i\rho)e^{-i(l+1/2)\phi}\end{matrix}\right),
\end{equation}
where $J_a$ is the Bessel function of the first kind of order $a$ (the second branch, with the modified Bessel function of the first kind $Y_a$, is ruled out as it diverges at $x=y=0$). Now the reduced radial equation becomes:
\begin{equation}
\label{diraceqcylred}
\left(\begin{matrix}\partial_zF\\
\partial_zG\end{matrix}\right)=\left(\begin{matrix}-\tilde{k} & \tilde{E}+\tilde{M}\\ \tilde{M}-\tilde{E} & \tilde{k}\end{matrix}\right)\left(\begin{matrix}F\\
G\end{matrix}\right)
\end{equation}
with $\tilde{k}=k/\sqrt{f}$ (let us note that Eq.~(\ref{diraceqcylred}) is for the pair $(F,G)$, whereas the initial equation (\ref{direqcyl}) is written for the bispinor $(FK_1,-GK_2)$). For the WKB calculation of the density, it is useful to remind that the wave function $\Psi$ in Eq.~(\ref{cylansatz}) has two quantum numbers corresponding to the motion in the $(x,y)$ plane: they are simply the momentum projections $k_x,k_y$ (or equivalently the momentum module $\lambda$ and the angular momentum $l$). The radial eigenfunctions in $z$-direction provide a third quantum number $n$.

\section{Equation of state of the WKB star}

In this section we construct the model of the bulk fermions in an improved semiclassical approximation -- the WKB star. We solve the Dirac equation in the WKB approximation, and the density is computed  by summing a large number of energy levels. This is in the spirit of Thomas-Fermi approximation. However, we perform an exact summation of a {\em finite} number of WKB quantum-mechanical solutions for the wave functions rather than approximating the sum by an integral as implied in the semiclassical fluid limit. One of the drawbacks of the Thomas-Fermi fluid limit are sharp bounds (i.e., discontinuous first derivative) of density and pressure profiles along the radial direction (see e.g. ~\cite{Hartnoll:es, Hartnoll:phtr,Larus}). As we have already argued, sharp bounds make it hard if not impossible to capture several phenomena. In this respect summing WKB wave functions goes beyond Thomas-Fermi; it includes quantum corrections as the number of occupied states is finite and all collective and individual profiles will be continuous without sharp edges. In further work one might
start from our model and treat the quantum-mechanical (one loop) corrections in a more systematic way in order to bridge the gap between the electron star~\cite{Hartnoll:es} and single-particle quantum mechanical calculation of Dirac hair~\cite{csz2010}.

\subsection{WKB hierarchy and semiclassical calculation of the density}

In the framework of quantum-many-body calculations, the first task is to construct the induced charge density $n(z)$. Physically, the origin of the induced charge in our model is the pair production in the strong electromagnetic field of the black hole. To remind the reader, a (negatively) charged black hole in AdS space is unstable at low temperatures, and spontaneously discharges into the vacuum \cite{Horowitz}. This means that there will be a non-zero net density of electrons $n(z)$. One can calculate $n(z)$ in a Hartree approximation as a density of non-interacting electrons, compute the collective effect on other fields by this density and iterate. Our novel approach is to use WKB methods to efficiently compute the many wave functions enumerated by the quantum numbers $(\lambda,l,n)$.

The algorithm for the WKB expansion of the wave function for Dirac equation is adopted from \cite{voskresenie}. Even though every single step is elementary, altogether it seems to be less well known than its Schr\"{o}dinger equivalent. We consider the Dirac equation in the form~(\ref{direqmat}) and introduce the usual WKB phase expansion:
\begin{equation}
\label{phasexp}\Psi(z)=e^{\int_{z_0}^zdzy(z)\sqrt{f(z)}}\chi(z)
\end{equation}
with the spinor part $\chi(z)$. The phase $y(z)$ can be expressed as the semiclassical expansion in $\hbar$ \footnote{From the very beginning we put $\hbar=1$. However, to elucidate the semiclassical nature of the expansion we give it here with explicit $\hbar$.
Dirac equation becomes $\hbar\sqrt{f}\partial_z\hat{\Psi}=\hat{D}\hat{\Psi}$, where $\hat{\Psi}=(\Psi_+,\Psi_-)$, yielding the expansion $y(z)=\hbar^{-1}\left(y_{-1}(z)+\hbar y_0(z)+\hbar^2y_1(z)+\ldots\right)$.}
\begin{equation}
\label{yexp}y(z)=y_{-1}(z)+y_0(z)+y_1(z)+\ldots
\end{equation}
The equations for the perturbative corrections now follow from (\ref{phasexp}-\ref{yexp}):
\begin{eqnarray}
\label{ym1}\hat{D}\chi_0&=&y_{-1}\chi_0,\\
\label{y0}\hat{D}\chi_1&=&y_{-1}\chi_1+y_0\chi_0+\sqrt{f}\partial_z\chi_0,\\
\ldots \nonumber\\
\label{yn}\hat{D}\chi_n&=&y_{-1}\chi_n+\sqrt{f}\partial_z\chi_{n-1}+\sum_{i=0}^{n-1}y_{n-i-1}\chi_i.
\end{eqnarray}
Notice in particular that $y_{-1}$/$\chi_0$ is an eigenvalue/eigenvector of $\hat{D}$. In our case the matrix $\hat{D}$ has rank two, so there are two eigenvalues/eigenvectors for $y_{-1}/\chi_0$: $y^\pm_{-1}$ and $\chi^\pm_0$. To find the first order correction to the phase of the wave function $y_0$, we multiply~(\ref{y0}) from the left by the left eigenvalue $\tilde{\chi}^{\pm}_0$ of the matrix $\hat{D}$ ($\hat{D}$ is in general not symmetric, so the right and left eigenvalues are different):
\begin{equation}
\label{wkbmom}y_0=-\frac{(\partial_z\chi^\pm_0,
\tilde{\chi}^\pm_0)}{(\tilde{\chi}^\pm_0,\chi^\pm_0)}.
\end{equation}
so we can now construct the usual WKB solution of the form $\Psi_\pm=e^{i\theta_\pm}/\sqrt{q}$, where $q$ is the WKB momentum and $\theta_\pm$ the phase. The term $y_0$ is just the first order correction to $\theta_\pm$.

Finally, let us recall the applicability criterion of the WKB calculation. It is known that WKB approximation fails in the vicinity of turning points. The condition of applicability comes from comparing the leading and the next to leading term in the expansion~(\ref{yexp}):
\begin{equation}
\frac{y_0(z)}{y_{-1}(z)}\ll 1.
\end{equation}
In terms of $\tilde{E}(z)$ and $\tilde{M}(z)$ introduced in Eq.~(\ref{mbdef}) it gives at $k=0$:
\begin{equation}
\label{wkbcrit}\frac{\tilde{M}(z)\partial_z\tilde{E}(z)-\tilde{E}(z)\partial_z\tilde{M}(z)}{\tilde{E}(z)(\tilde{E}(z)-\tilde{M}(z))}\ll 1.
\end{equation}

\subsubsection{WKB wave function}
According to~(\ref{ym1}), the leading effective WKB momentum for the motion in $z$ direction $q \equiv \vert y^{\pm}_{-1}\vert$ is:
\begin{equation}
\label{cyly}q^2(z)=\tilde{E}^2(z)-\tilde{M}^2(z)-\tilde{k}^2(z).
\end{equation}
The wave function in radial direction, $\Psi=(F,-G)$, is given by the superposition of two linear independent solutions
\begin{equation}
\label{dirsol}
\Psi(z)=C_+\chi_+(z)e^{i\theta\left(z\right)}+C_-\chi_-(z)e^{-i\theta\left(z\right)},
\end{equation}
with the phase determined by
\begin{eqnarray}
\theta(z)&=&\int^z\left(q(z')+\delta\theta(z')\right)dz'\\
\delta\theta(z)&=&\int^z\frac{\tilde{k}\partial_z\tilde{k}-q\partial_zq+\left(\tilde{E}-\tilde{M}\right)\left(\partial_z\tilde{E}+\partial_z\tilde{M}\right)}{2\tilde{k} q}dz.
\end{eqnarray}
The constants $C_+$ and $C_-$ are related by invoking the textbook boundary conditions \cite{ll3} for the behavior of WKB wave function at the boundary of the classically allowed region ($q^2(z)>0$) and the classically forbidden region ($q^2(z)<0$). The wave function in the classically allowed region then reads:
\begin{eqnarray}
\label{cylfun}
\Psi(z)&=&\frac{C}{\sqrt{q(z)}}\left(\begin{matrix}\sqrt{\tilde{E}(z)+\tilde{M}(z)}\sin\left(\theta(z)-\delta\theta(z)\right)\\ \sqrt{\tilde{E}(z)-\tilde{M}(z)}\sin\theta\left(z\right)\end{matrix}\right),\\
\delta\theta(z)&=&\mathrm{ArcSin}\frac{q(z)}{\sqrt{\tilde{E}^2(z)-\tilde{M}^2(z)}},
\end{eqnarray}
and $C$ is the only remaining undetermined normalization constant. Integrating the probability density over all coordinates in the classically allowed region $(z_1,z_2)$ gives the normalization condition:
\begin{equation}
C^2\int_0^1dz\frac{\sqrt{g_{3d}(z)}}{a(z)^2}\int dx\int dy C_{2d}^2\Psi_{n k_xk_y}(z,x,y)\Psi_{n'k_x'k_y'}^\dagger
(z,x,y)=1.
\end{equation}
The metric factor is $g_{3d}(z)=g(z)g^{tt}(z)$, and $a(z)$ is the conversion factor from~(\ref{chitrans}). In the left-hand side of the equality we took into account the normalization of the continuous spectrum in the $(x,y)$ plane. The integration in the perpendiular coordinates is trivial for the solution (\ref{ksol}), as we can transform the integral into the integral over $\rho,\phi$ and the orthogonality relation for Bessel functions gives the definition of $C_{2d}^{2}$:
\begin{equation}
\label{normbessel}
C_{2d}^{-2}\int_0^\infty J(\lambda\rho)J(\lambda'\rho)\rho d\rho=\frac{\delta(\lambda-\lambda')}{\lambda}
\end{equation}
and it allows us to express the normalization constant as:
\begin{equation}
\label{const}C=\left(4\pi\int dz\frac{\sqrt{g^{tt}}}{\sqrt{g^{zz}}}\frac{\tilde{E}(z)}{q(z)}\right)^{-1/2},
\end{equation}
where a factor of $2\pi$ comes from the integration over $\phi$ and an additional factor of $2$ from the summation over the full four-component wave function, i.e. bispinor (each spinor gives $\tilde{E}(z)/q(z)$ after averaging over the fast oscillating phase $\theta$). This completes the derivation of WKB wave function and allows us to compute the density.

\subsubsection{WKB density}

As in \cite{voskresenie} we find the total density by summing single-particle wave functions in the classically allowed region. The WKB wave function is characterized by the quantum numbers $(\lambda,l,n)$ with $\lambda$ being the linear momentum in the $x-y$ plane, $l$ -- the orbital momentum in the $x-y$ plane and $n$ -- the energy level of the central motion in the potential well along $z$ direction. The bulk density can be expressed as the sum over the cylindrical shells of the bulk Fermi surface. Each shell satisfies the Luttinger theorem in the transverse ($x-y$) direction and so the density carried by each shell $n_{xy}(z)$ can easily be found. We can then sum over all shells to arrive at the final answer which reads simply $\int dzn_z(z)n_{xy}(z)$. A similar qualitative logic for summing the Luttinger densities in the $x-y$ plane was used also in \cite{Sachdev} although the model used in that paper is overall very different (see also the fully consistent treatment with regularization in \cite{newref}).

Let us start by noticing that the end points of the classically allowed region determine the limits of summation over $n$ and $\lambda$: $q^2(\omega_n,\lambda)\ge 0$. Thus, the density in the WKB region is:
\begin{equation}
n(z)=\frac{2\pi}{a(z)^2}\int_0^{2\pi}d\phi\sum_{n:q^2(\omega_n,\lambda)\ge 0}\int_0^{\sqrt{f(z)(\tilde{E}^2(\omega,z)-\tilde{M}^2(z))}}\lambda d\lambda\int_0^\infty d\rho\rho C_{2d}^2|\Psi(z,x,y)|^2.
\end{equation}
The limit of the sum over the level number $n$ is determined by the requirement that WKB momentum be positive; in other words, we sum over occupied levels inside the potential well only. Remember that the bulk fields live at zero temperature, hence there is no Fermi-Dirac factor. The sum over the orbital quantum number $l$ extends to infinity as the $(x,y)$ plane is homogenous and the orbital number does not couple to the non-trivial dynamics along the radial direction. We can now invoke the (local) Bohr-Sommerfeld quantization rule:
\begin{equation}
\label{bohr}\int dzq(z)=N_{WKB}\pi
\end{equation}
to estimate the total number $N_{WKB}$ of radial harmonics in the sum. The expression for $N_{WKB}$ in combination with (\ref{const}) then give:
\begin{equation}
C_n=\left(\frac{1}{4\pi^2}\frac{\partial\omega_n}{\partial n}\right)^{1/2}, \text{ for }q(z)\gg\delta\theta(z), \text{ } z\approx 1.
\end{equation}
Now we turn the summation over the quantum number $n$ into the integration over energy and obtain for the \emph{bulk} electron density (here we also performed the integration over $\rho$ using the explicit expression for the wave function~(\ref{direqcyl}) and the normalization condition (\ref{normbessel}) for the Bessel functions):
\begin{equation}
\label{nz11}
n(z)=\frac{4\pi}{a(z)^2}\int_0^{2\pi}d\phi\int_0^{\sqrt{f(z)(\tilde{E}^2(0,z)-\tilde{M}^2(z))}}d\lambda\lambda\int_0^{\mu_{loc}} d\omega\frac{\tilde{E}(\omega,z)}{4\pi^2q(\omega,\lambda,z)}.
\end{equation}
After performing first integral over $\omega$ and then over $\lambda$ we get\footnote{The given result for $n$ can be compared to the charge density in the electron star limit given in \cite{Hartnoll:2011}. The metric functions used there are related to ours as $f\mapsto fe^{-h}/z^2$ and $g\mapsto 1/fz^2$, where our metric functions are on the right hand side. Likewise, our definition of $q_{WKB}$ is related to $k_F$ of \cite{Hartnoll:2011} as $q_{WKB}=k_F/\sqrt{f}$. Now the \emph{total} bulk charge is expressed in \cite{Hartnoll:2011} as $Q=\int dz\tilde{n}_e(z)$ where $\tilde{n}_e(z)\sim n(z)e^{h/2}$. In our conventions $Q=\int dz\sqrt{-g}g^{zz}g^{tt}n=\int dzn(z)e^{h/2}$ thus giving the same result as in \cite{Hartnoll:2011}.}:
\begin{equation}
\label{WKBdens}n(z)=z^3\frac{q_{WKB}^3f^{3/2}(z)}{3\pi^2}
\end{equation}
with $q_{WKB}$ determined by
\begin{equation}
q_{WKB}^2=\tilde{E}^2(0,z)-\tilde{M}^2(z)\label{pmax}.
\end{equation}
Notice that this formula corresponds with common knowledge on the density of electron star \cite{Hartnoll:es}. However, even though the formal expression is the same, the self-consistent solution for the metric and gauge field is different because of the quantum correction we introduce to pressure. The difference is visualized in Fig.~1A where we preview our backreacted WKB star solutions and compare them to the semi-classical (electron star) limit. While the electron star density exhibits a discontinuity at the horizon, the WKB density smoothly falls off to zero. However, both models have a semiclassical "edge": outside the region $z_1<z<z_2$, the density is exactly zero. In reality, quantum tails change this picture. In \cite{thesis} we show that (small) nonzero density extends all the way between the boundary and the horizon. However, it is not expected to change the finite temperature physics which is in the focus of this paper. We therefore do not take into account the quantum tails in further calculations, to avoid any distractions from the main message.
%tilde{M}^2(z)-3\left(\partial_z\log\left(\tilde{E}(z)+\tilde{M}(z)\right)\right)^2+\frac{1}{2}\frac{\partial_{zz}(\tilde{E}(z)+\tilde{M}(z))}{\tilde{E}(z)+\tilde{M}(z)},

\begin{figure}[ht!]
\begin{center}
(A)\includegraphics[width=0.45\linewidth]{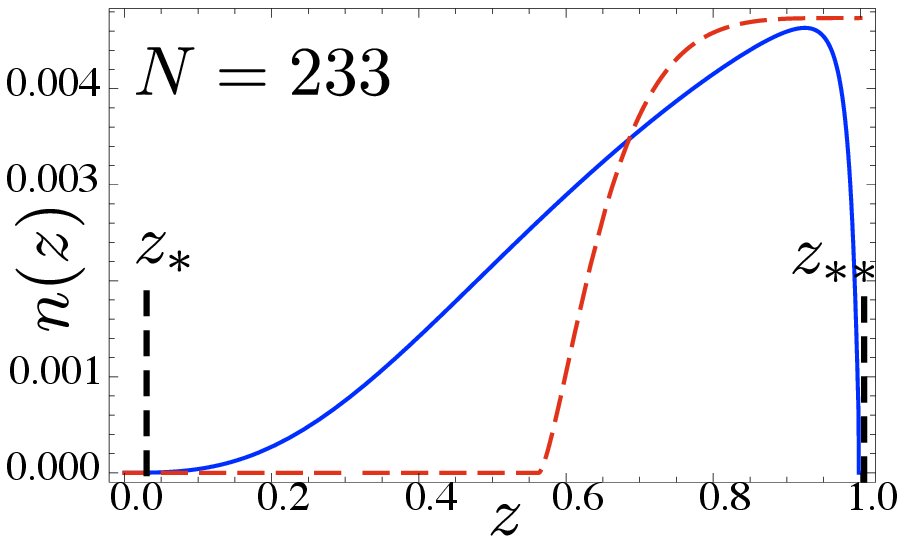}
(B)\includegraphics[width=0.45\linewidth]{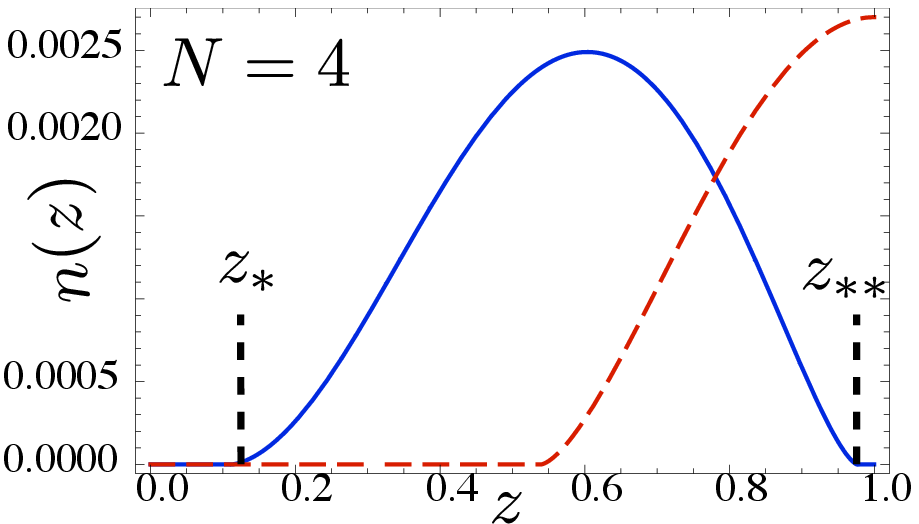}
(C)\includegraphics[width=0.45\linewidth]{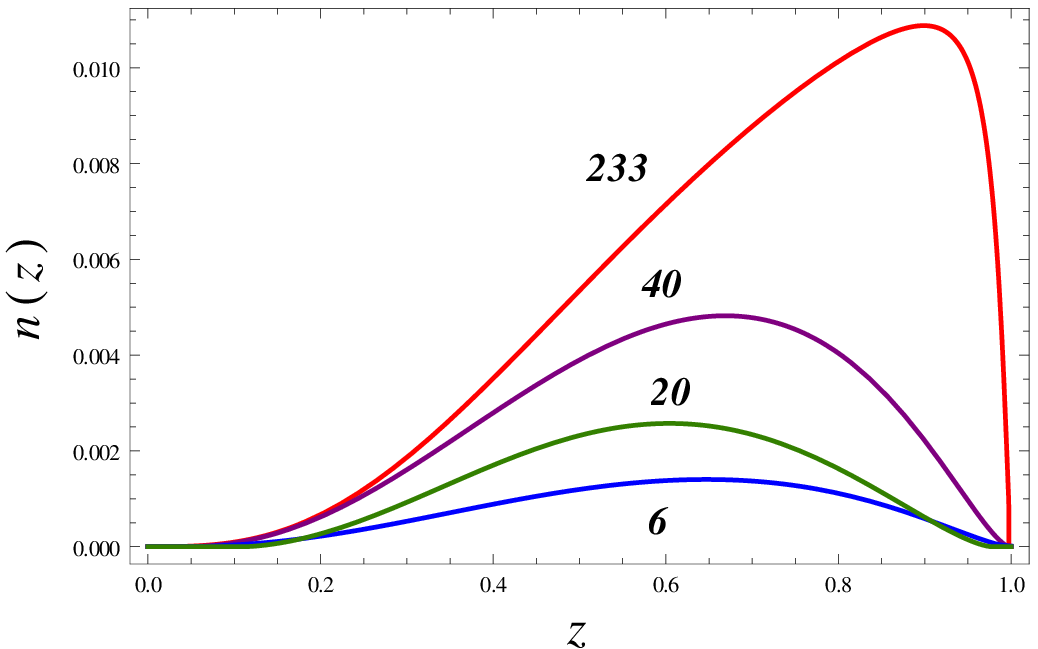}
(D)\includegraphics[width=0.45\linewidth]{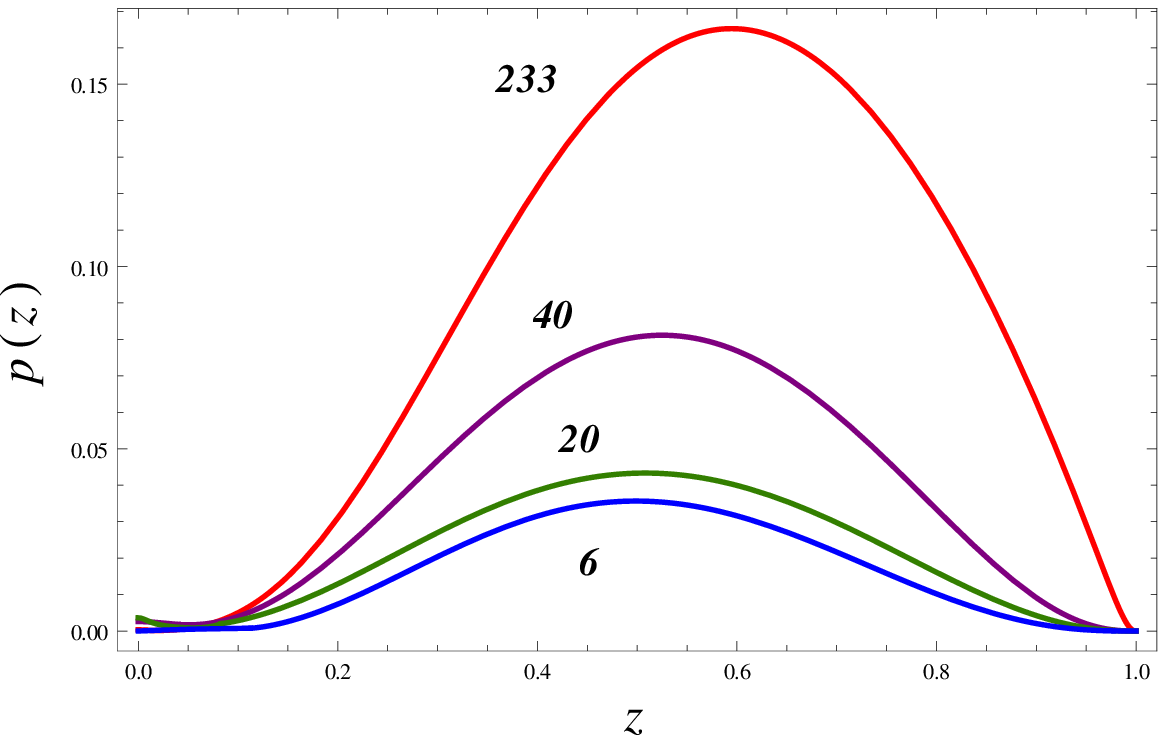}
\caption{WKB bulk density $n(z)$ (Eq.~\ref{WKBdens}, blue lines) and electron star density (red dashed lines). Parameter values (A) $(\mu,e,m)=(1.7,1,0.1)$, (B) $(\mu,e,m)=(1.7,10,1)$. The classically allowed region lies between the turning points $z_*$ and $z_{**}$, determined by the condition of vanishing WKB momentum ($q(z_*)=q(z_{**})=0$). The parameters for (A) are in the classical (electron star) regime, with $N_{WKB}\gg 1$ when WKB approximation is quite accurate. The plot (B) shows a case of small $N_{WKB}$ where the WKB approximation becomes inadequate and further quantum corrections are likely to be important. (C) Bulk density for a range of values $(\mu,e,m)=(1.7,1,0.1)$ (red), $(\mu,e,m)=(1.7,5,0.1)$ (violet), $(\mu,e,m)=(1.7,10,1)$ (green) and $(\mu,e,m)=(1.7,20,1)$ (blue). For large specific charge of the fermion (and therefore a large number of WKB levels in the bulk) the solution is dominated by the classically allowed region and looks similar to the electron star limit. For smaller $e/m$ values (and thus fewer WKB levels) the quantum correction in the near-boundary region becomes more important and the curves are visibly different from the fluid limit. (D) Thermodynamical pressure (Eq.~\ref{pclass0}), for the same parameter values as in (C).}
\label{figpress3}
\end{center}
\end{figure}

\subsection{Pressure and equation of state in the semiclassical approximation}
\label{secpress}

Following the logic behind the density calculation, we will now calculate the pressure $p$ along the radial direction. It will actually prove easier to derive the expression for the (bulk) internal energy density first and then calculate the pressure. By definition, the energy density reads
\begin{equation}
\label{energy}\mathcal{E}(z)=\sum_{k_x,k_y}\int dx\int dy\int_0^{\mu_{loc}}d\omega\omega\Psi^\dagger(z)\Psi(z)=\sum_\lambda\int_0^{\mu_{loc}}d\omega\omega \frac{\tilde{E}(z)}{4\pi^2q(z)}
\end{equation}
where $\tilde{E}(z)$ is defined in (\ref{mbdef}), $\mu_{loc}=\mu e^{h(z)/2}/f(z)$ and the sum limits are the same as in \eqref{nz11}. Performing the integration in a similar fashion as when computing $n(z)$ in (\ref{nz11}-\ref{WKBdens}), we obtain
\begin{equation}
\label{energ}\mathcal{E}=\frac{3}{4}e\Phi
n+\frac{1}{4}f^2\tilde{M}^2\mathrm{ArcSinh}\frac{\tilde{E}}{\tilde{M}}.
\end{equation}
Notice that the first term exactly corresponds to the electrostatic energy while the second is the one-loop term that encapsulates the quantum fluctuations. The above result is remarkably close to the Hartree vacuum polarization correction as it appears in various model energy functionals in literature.

\begin{figure}[ht!]
\begin{center}
(A)\includegraphics[width=0.45\linewidth]{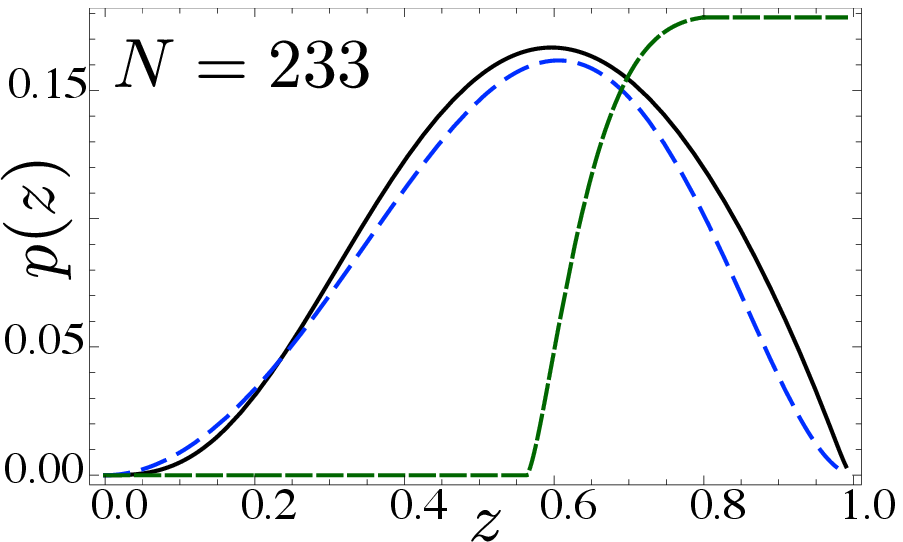}
(B)\includegraphics[width=0.45\linewidth]{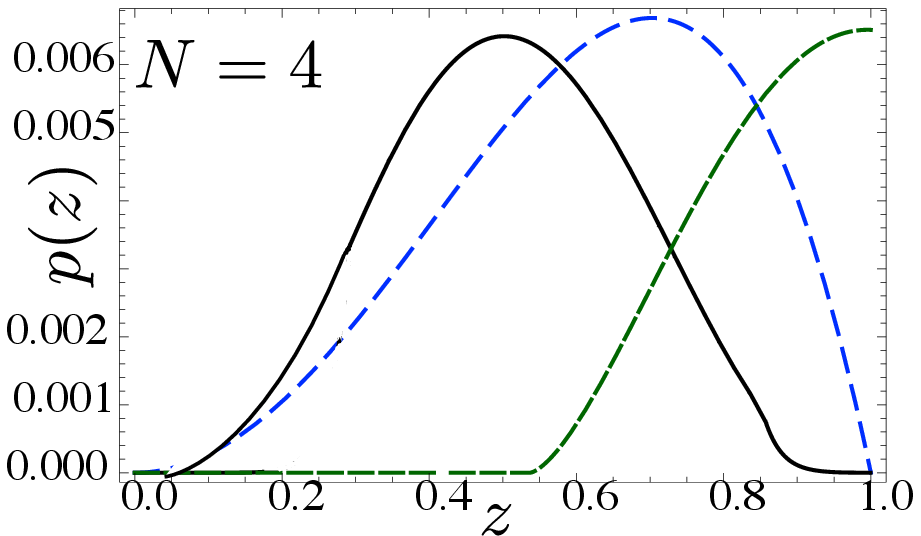}
\caption{Comparison between full quantum pressure (dashed blue lines, Eq.~\ref{press3}) and thermodynamic pressure (solid black lines, Eq.~\ref{pclass0}) for two sets of parameters: $(\mu/T,e,m)=(1.7,1,0.1)$ (A) and $(\mu/T,e,m)=(1.7,5,1)$ (B). For comparison we plot also the fluid pressure $p=en\Phi/2$ (dashed green lines). Expectedly, for $N_{WKB}\gg 1$ (A) the thermodynamic approximation comes close to the exact summation while for $N_{WKB}$ small the level spacing is large and the thermodynamic limit is no longer a good approximation to the sum of the contributions of individual levels. Notice that both ways of computing pressure yield similar results for large $N_{WKB}$ but deviate at smaller $N_{WKB}$.}
\label{figpress}
\end{center}
\end{figure}

\subsubsection{Microscopic pressure}

The easiest way to express the pressure is to make use of the first law of thermodynamics, which states
\begin{equation}
\label{firstlaw}p(z)=\sqrt{g^{zz}}\left(e\Phi\left(z\right)-\mathcal{E}\left(z\right)\right).
\end{equation}
There are two possible approaches to arrive at the pressure also directly from the equations of motion. We can express the radial pressure $p$ from the microscopic fermionic Lagrangian (\ref{lagfermi}). By definition it reads
\begin{equation}
\label{press30}p=\sum_{n,\lambda}\left(\Psi_+^\dagger\sigma_3\partial_z\Psi_-+\Psi_-^\dagger\sigma_3\partial_z\Psi_+\right)=\frac{1}{a^2}\left(\tilde{E}(F^2+G^2)-\tilde{M}(F^2-G^2)-2\tilde{k}FG\right)
\end{equation}
The equality follows directly from the Dirac equation, substituting the expressions for $\partial_z\Psi_\pm$ from (\ref{direqmat}). Now we can exploit the lowest order WKB solution (\ref{cylfun}) to get
\begin{equation}
\label{press31}p=\frac{2\pi}{a^2}\sum_{n,\lambda}C_n^2\left(\frac{\tilde{E}-\tilde{M}}{q}-\tilde{k}\right),
\end{equation}
which, after the momentum integration, gives:
\begin{equation}
\label{press3}p=2\pi\sum_nC_n^2e^{h/2}z^3\sqrt{f}\left[\left(\tilde{E}-\tilde{M}\right)q_{WKB}^2(z)-\frac{2}{3}q_{WKB}^3(z)\right]
\end{equation}
The explicit calculation is tedious but straightforward. Unlike the density case, the final sum is not readily performed to obtain a closed-form expression. Instead, we integrate numerically over the energy levels $\omega_n$ to obtain the function $p(z)$. However, even a quick look at (\ref{press3}) tells that it behaves as $q_{WKB}^3$ at leading order, for $q_{WKB}$ large (the first and the third term will contribute as $q_{WKB}^3$). After the energy integration this term gains roughly a factor of $\mu$, implying that $p\sim\mu n\sim\mu^4$, as we expect to recover in the fluid limit.

We have now calculated the radial pressure, i.e. the fermionic component of the stress tensor $T_z^z$. Due to local isotropy, it does not depend on the direction and position in the $x-y$ plane. The same happens in the fluid limit, as shown in \cite{Hartnoll:es}. The pressure in the perpendicular direction (in the $x-y$ plane) is analogously expressed as
\begin{equation}
\label{pressxy0}p_\perp=-\sum_{n,\lambda}ik\left(\Psi_+^\dagger\sigma_1\Psi_-+\Psi_-^\dagger\sigma_1\Psi_+\right)=
\frac{2\pi}{a^2(z)}\sum_{n,\lambda}C_n^2\frac{1}{q}\lambda\tilde{E}
\end{equation}
The summation over $\lambda$, i.e. the value of the in-plane momentum can again be performed analytically, yielding:
\begin{equation}
\label{pressxy}p_\perp=2\pi\sum_nC_n^2e^{h/2}z^3fq_{WKB}^2\tilde{E}.
\end{equation}
In fact, the above sum has a closed-form limit for $N_{WKB}\to\infty$:
\begin{equation}
\label{pressxy1}p_\perp=f^2e^{h/2}z^3\frac{q_{WKB}^4}{12\pi^2},
\end{equation}
which obeys the relation $p_\perp\sqrt{g_{ii}}=\tilde{\mu}n\sqrt{g_{00}}/4$, the covariant version of the relation $p=\mu n/4$. We will not make use of $p_\perp$ as the $ii$ component of the Einstein equations is not functionally independent of the $00$ and $zz$ components; the two metric functions $f,h$ are determined from the two equations, and the third one can only serve as a consistency check.

\subsubsection{Thermodynamic pressure}

In a ``near''-classical regime, at large occupation number, thermodynamics ought to work, so we can express the pressure from the energy density $\mathcal{E}$, as $p=-\partial E/\partial V$. This expression is still hard to calculate exactly. However, we can use the following trick to estimate $p$ at the leading order. Consider a small change of the number density $\delta n$. It will introduce a small change of energy $\delta\mathcal{E}$, pressure $\delta p$ and the volume of the bulk electron gas $\delta V$, the latter because the classically allowed region where $q_{WKB}^2>0$ will shift and grow (if $\delta n>0$). Now since the metric is radially symmetric we can expand the volume $V=\int_{z_*}^{z_{**}} d^3x\frac{e^{-h(z)/2}}{z^4}$ around its initial value and find that the leading term in its variation behaves as $\delta V=V\delta\ell/(1-\ell)+\ldots$, where $\ell\equiv z_{**}-z_*$ is the (dimensionless) length of the classically allowed interval along the $z$ axis, i.e. the interval between the zeros of the WKB momentum $q_{WKB}(z)=\sqrt{\tilde{E}^2(z)-\tilde{M}^2(z)}$. This yields
\begin{equation}
\frac{\partial E}{\partial V}=\mathcal{E}+V\frac{\partial\mathcal{E}}{\partial V}=\mathcal{E}+V\frac{\delta\mathcal{E}(1-\ell)}{V\delta\ell}=\frac{\delta\mathcal{E}}{\delta\ell}.
\end{equation}
Since all the processes we study are certainly adiabatic (looking at the whole system of gravity plus the matter fields), we can replace the variations by partial derivatives and write $p\sim\partial\mathcal{E}/\partial\ell$ as an approximation for the radial pressure. However, even this expression we are only able to evaluate in a very crude way. For $N_{WKB}\gg 1$, it is natural to assume (and confirmed by the numerics, see Fig.~1) that $z_{**}$ is very close to the horizon, $z_*$ is quite far from the horizon and $\ell\approx 1-z_*$. For $z\sim z_*$, we assume that the electric potential does not deviate much from the linear law: $\Phi\sim\mu (1-z)$, because $z_*$ is not far from the boundary. This means that the metric function $h(z)$ can be well approximated by a linear function $h(z)\sim\mathrm{const.}(1-z)$. Solving the equation $q_{WKB}^2=\tilde{E}^2(z_*)-\tilde{M}^2(z_*)=0$, we get $\ell\sim 1-\log\frac{e^2\mu^2}{m^2}$, and (\ref{energ}) gives the thermodynamic pressure. However, we cannot get the numerical prefactor right in our approach, and this is important in order to satisfy the first law of thermodynamics, which in the fluid limit predicts $p=\mathcal{E}/4$. We therefore norm $p_{thd}$ by hand by a constant factor $C_{thd}$. This gives:
\begin{equation}
\label{pclass0}p_{thd}=-C_{thd}\frac{\partial\mathcal{E}}{\partial\Phi}\frac{\partial\Phi}{\partial\mu}\frac{\partial\mu}{\partial\ell}\sim\frac{1}{4}e\mu(1-z)\left(n+\frac{\tilde{M}^2e^{-h}}{z\sqrt{\tilde{M}^2+e^h\tilde{E}^2}}\right)
\end{equation}
This is the relevant regime to compare with the electron star. We will call the estimate (\ref{pclass0}) thermodynamic pressure and denote it by $p_{thd}$ to differentiate from the exact summation of WKB wave functions (\ref{press3}). These expressions are also the equations of state of the system as they connect the pressure to the density. The thermodynamic pressure is more convenient for calculations. In spite of its approximate nature, (\ref{pclass0}) in particular yields a remarkably accurate result when compared with the quantum pressure at $N_{WKB}\gg 1$.

We can make the connection between the exact first law of thermodynamics (\ref{firstlaw}) and the quick estimate (\ref{pclass0}) by showing them to be equal in the limit of small $\tilde{E}$, which is appropriate in the vicinity of the phase transition from WKB star to the RN black hole: since $\tilde{E}\sim\mu/T$, this really means high temperature regime, $T\ll\mu$. In this case expanding both equations in $\tilde{E}$, we find the same expression:
\begin{equation}
\label{pexpand}p\approx\frac{1}{4}e\Phi n+\frac{1}{4}\frac{f}{z}\tilde{E}\tilde{M}+O(\tilde{E}^3).
\end{equation}
Finally, it is illustrative to see how we reproduce the electron star pressure \cite{Hartnoll:es} in the limit of large density. For $n\to\infty$, the first term in $\mathcal{E}$ and $p_{thd}$ dominates and we obtain from (\ref{energ}) and (\ref{pclass0})
\begin{equation}
\label{pclass}p_{ES}=\frac{1}{4}e\Phi n
\end{equation}
as expected for an ideal fluid, which corresponds to the electron star approach. The physical interpretation of this result (and of the pressure inside the classically allowed region in general) is that of a Fermi gas pressure which, as we know, survives also in the limit of classical thermodynamics. The comparison of $p$, $p_{thd}$ and $p_{ES}$ is summarized in Fig. \ref{figpress}, for high and low number of levels. While the thermodynamic approximation (\ref{pclass0}) is good when $N_{WKB}\gg 1$, for small $N_{WKB}$ both the fluid limit and the thermodynamic limit eventually break down and the contributions of individual levels must be taken into account. Once again, the introduction of Airy corrections would extend the nonzero pressure to the whole AdS space, which is only expected to be relevant at $T=0$ \cite{thesis}.

\section{Maxwell-Dirac-Einstein system}

We have now arrived at the point where we can look for a numerically self-consistent solution of the Einstein-Maxwell equations. The numerics uses an iterative procedure to converge toward the solution. Only in the IR region it is possible to use a scaling ansatz to estimate the scaling behavior of the metric and matter fields, akin to the procedure used in \cite{Review:2010}. This is the first attempt at a numerically self-consistent solution including backreaction on the geometry with holographic fermions which goes beyond the fluid picture of \cite{Hartnoll:es}.

Our calculation is similar to the one for relativistic ideal fluid (i.e. electron star) approximation. Because an ideal fluid is dissipation-less one can construct an action as put forward in \cite{ActionFluid} and used in \cite{Hartnoll:es, Hartnoll:phtr}. The Lagrangian of this charged fluid coupled to gravity and electromagnetism is
\begin{equation}
\label{sfluid}S=\int d^4x\left[\frac{1}{2\kappa^2}\left(R+6\right)-\frac{1}{2q^2}\left(\partial_z\Phi\right)^2+p\right].
\end{equation}
In other words, the contribution of fermions reduces to the pressure $p$. While we do not take the fluid limit in this paper, within the WKB star model we assume that in the first approximation the influence of the corrections to fluid limit ($N_{WKB}\to\infty$) is fully encapsulated by the correction to the classical (or fluid) pressure we found in (\ref{energ}-\ref{press3}). The emergent isotropy and its implied ideal nature of the fluid at large occupation number should ensure this.

To construct the backreacted geometry, we therefore ``replace'' the fermionic terms in the exact action (\ref{action}) with our effective ideal fluid model in terms of the density and pressure of the bulk fermions. The total effective action is represented as $S=S_E+S_M+S_f$, the sum of Einstein, Maxwell and fluid part. The only nonzero component of the gauge field is $\Phi$ and the only non-vanishing derivatives are the radial derivatives $\partial_z$ (the others average out to zero for symmetry reasons). The nonzero fermion pressure $p$ is that considered in Sec. \ref{secpress} and there is a nonzero (local) charge density
\begin{equation}
\label{ndens}j^0_e=qn\sqrt{g^{00}}=qn\frac{ze^{h/2}}{\sqrt{f}}.
\end{equation}
The fermion fluid term in the effective action thus becomes
\begin{equation}
\label{fermiaction}
S_f=-\int d^4x\sqrt{-g}\left(j^0_e\Phi+p\right).
\end{equation}
Due to the preserved spherical symmetry we may substitute these simplifications directly in the effective action to  arrive at:
\begin{equation}
\label{allaction}
S_{eff}=\int d^4x\sqrt{-g}\left[\frac{1}{2\kappa^2}\left(R+6\right)-\frac{z^4}{2}e^h\left(\frac{\partial\Phi}{\partial z}\right)^2-j^0_e\Phi+p\right].
\end{equation}
The only components of the stress tensor the fermion kinetic energy contributes to are the diagonal ones; the others vanish due to homogeneity and isotropy in time and in the $x-y$ plane. From (\ref{allaction}) we get the equations for the energy-momentum tensor:
\begin{eqnarray}
\label{enmom00}T_0^0 &=&-\frac{1}{2}z^4e^h\left(\frac{\partial
\Phi}{\partial z}\right)^2+j^0_e\Phi\\
\label{enmomzz}T_z^z=&=&-\frac{1}{2}z^4e^h\left(\frac{\partial
\Phi}{\partial z}\right)^2+j^0_e\Phi+mn+g_z^zp.
\end{eqnarray}
With the metric ansatz (\ref{metric}), we can now write down our equations of motion:
\begin{eqnarray}
\label{tfeins1}\frac{1}{\sqrt{-g}}\left(\partial_ze^{-h/2}\partial_z\Phi\right)&=&-j_e^0\\
\label{tfeins2}3f-z\partial_zf-3&=&T_0^0\\
\label{tfeins3}3f-z\partial_zf-3zf\partial_zh-3&=&T_z^z.
\end{eqnarray}
Notice that the $ii$ component of the Einstein equations:
\begin{equation}
\label{tfeinsii}\frac{1}{2}\left(\partial_{zz}f-f\partial_{zz}h\right)+\frac{1}{4}\left(f\partial_zh^2-3\partial_zf\partial_zh\right)+\frac{f\partial_zh-2\partial_zf}{z}+\frac{3f}{z^2}=T_i^i
\end{equation}
with
\begin{equation}
\label{tii}T_i^i=-\frac{1}{2}z^4e^h\left(\frac{\partial
\Phi}{\partial z}\right)^2+g_i^ip_\perp
\end{equation}
is functionally dependent on the others and drops out. For that reason, (\ref{tfeins1}-\ref{tfeins3}) forms the complete system of Maxwell-Einstein equations. We do \emph{not} need to know $T_i^i$ or $p_\perp$ nor to assume the isotropy (in the sense $T_i^i=T_z^z$).

In this article we shall only be interested in finite temperature solutions. The gravitational background is therefore a black hole with an horizon: a single zero in the warp function $f(z)$ at a finite value $z=z_H$.\footnote{At zero temperature, when the horizon vanishes due to fermionic backreaction (this includes also the case of Lifshitz geometry), the boundary condition for $f$ guarantees also the smoothness of the solution on the horizon: $\partial_zf(z_H)=0$. This condition ensures that we pick the correct branch of the solution as there are typically two families of functions $f(z)$ that satisfy the equations of motion and the condition $f(z)=0$. One of them has a vanishing derivative whereas the other has finite derivative as $z\to 1$.} Physically the inescapability of the black hole horizon immediately suggests the following boundary conditions. The black hole horizon should have no hair so $\Phi(z_H)=0$; $h(z)$ which characterizes the ratio of the UV and IR speed of light should be finite at the horizon: $h(z_H)=h_0$. Note that the effective WKB potential felt by the fermions blows up at the horizon and that the fermion wavefunctions therefore manifestly vanish at $z_H$. This same phenomenon is noted in the electron star at finite temperature which also has an ``inner'' edge outside the horizon \cite{Hartnoll:phtr,Larus}.

At AdS infinity the boundary conditions are standard in AdS/CFT: for the gauge field $\lim_{z\rightarrow 0} \Phi(z)=\mu$ fixes the chemical potential at the boundary ($z_0\to 0$). We normalize $\lim_{z\rightarrow 0} f(z)=1$, $\lim_{z\rightarrow 0} h(z)=0$. Again the boundedness of the normalized WKB wavefunctions uniquely fixes the behavior of the fermions.

Finally, it remains to define the units used throughout the paper. The natural unit of energy and momentum is the chemical potential $\mu$ and we will express all quantities in units of $\mu$. The two thermodynamic parameters are the chemical potential $\mu$ and $T$. As AdS/CFT is built on conformal field theories which have no intrinsic scale, the physics only depends on the ratio $\mu/T$.

Let us conclude with an outline of the numerical algorithm, which is not completely trivial. The boundary conditions to be implemented are given at different points: some are given at the AdS boundary and some at the horizon. Since the system is nonlinear, it is necessary to either linearize the system or to shoot for the correct boundary conditions with the full nonlinear system. After experimenting with both, we have decided to iterate the full, non-simplified system of equations, integrating from the horizon and shooting for the conditions at the boundary. The iterative procedure consists of two steps: we start with the non-backreacted AdS-RN geometry and compute the density (semiclassical plus the quantum corrections) for the the electron charge equal to $e/N$ (where $e$ is the physical charge and $N$ some positive integer), then we solve the system of Einstein-Maxwell equations (\ref{tfeins1}-\ref{tfeins3}), afterwards we increase the fermion charge to $2e/N$, calculate the charge density in the background $(f,h,\Phi)$ taken from previous iteration and solve for this density the Einstein-Maxwell equations (\ref{tfeins1}-\ref{tfeins3}). We repeat this procedure for charge $3e/N$, $4e/N$ etc.  After $N$ iterations we have arrived at the physical value of the charge $e$. Then we do more iterations with fixed charge $e$ to ensure that the solution has converged, checking that the set of functions $(f,h,\Phi)$ does not change from iteration to iteration. In this way we achieve the self-consistent numerical solution of the Maxwell-Dirac-Einstein system of equations. The integration is always done from the horizon, shooting for the conditions for $\Phi$ and $h$ at the boundary, since it is well known that integrating from the AdS boundary is a risky procedure as it is next to impossible to arrive at the correct branch of the solution at the horizon.

\section{Phases of holographic fermions}

\begin{figure}[ht!]
\begin{center}
(A)\includegraphics[width=0.45\linewidth]{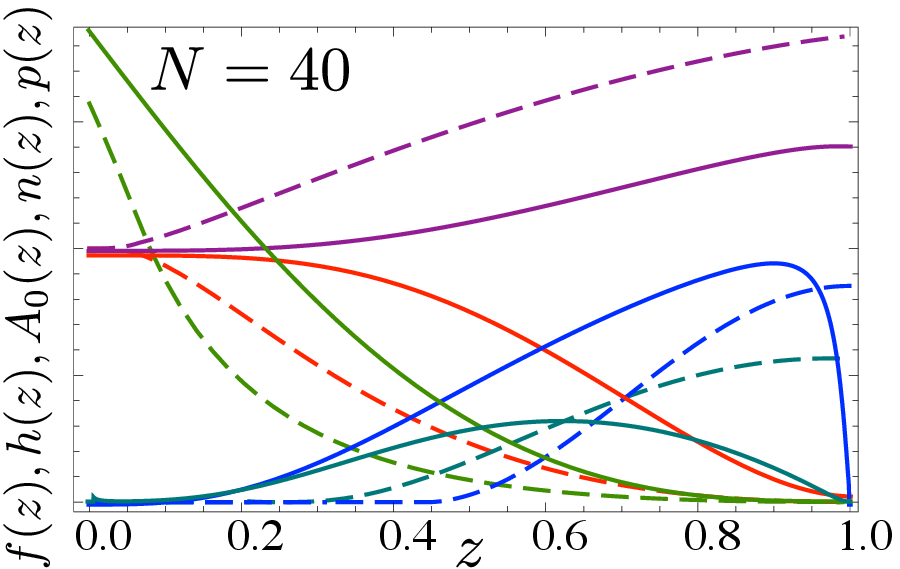}
(B)\includegraphics[width=0.45\linewidth]{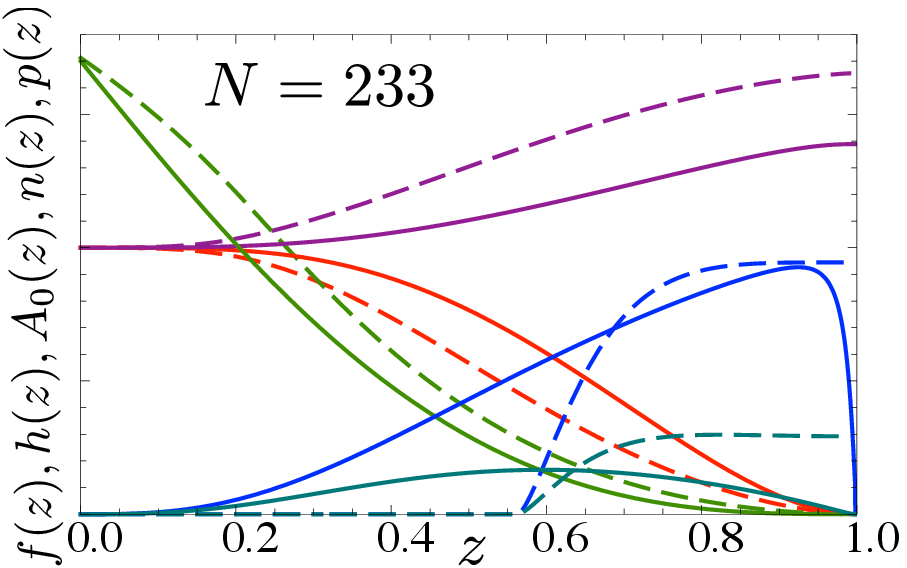}
\caption{Profiles of the metric functions $f(z)$ (red) and $e^{-h(z)}$ (violet), the gauge field $\Phi(z)$ (green), density $n(z)$ (blue) and the pressure $p(z)$ (cyan) at zero temperature, for $(\mu/T,e,m)=(1.7,1,0.1)$ (A) and for $(\mu/T,e,m)=(1.7,10,0.1)$ (B). Solid lines are calculated from our model while dashed lines are the electron star solution for the same parameter values. For better visibility density and pressure are rescaled by a constant factor. Near the boundary we always have $h(z)\to 0$ and $\Phi(z)=\mu+O(z)$, in accordance with the universal AdS asymptotics of the solution but in the interior the solutions start to deviate. Most striking is the absence of sharp classical edges in density and pressure. The difference in pressure will turn out to be crucial in moving away from the fluid limit. Here we have not shown the solution with $N_{WKB}=4$: this case deviates from the electron star ($N_{WKB}\to\infty$) so strongly that it does not make sense to compare it. Indeed, $4\ll\infty$!}
\label{figsol}
\end{center}
\end{figure}

We can now analyze the structure of both the bulk and the field theory side as a function of the parameters $T/\mu$, $e$ and $m$. We first shortly discuss the nature of the bulk solution for the geometry and gauge field and notice some qualitative properties. The typical way that the solutions to the WKB-Fermi-Einstein system (\ref{tfeins1}-\ref{tfeins3}) look is illustrated in Fig. \ref{figsol}. The near-horizon scaling of the metric and gauge field is of Lifshitz type, as expected in the light of earlier models \cite{polchin,Hartnoll:es}. Notice that we are working at finite temperature and thus do \emph{not} impose the IR boundary conditions for the metric functions which correspond to the Lifshitz geometry. Our finding of Lifshitz scaling is purely numerical, with the simple boundary conditions discussed above. In the figure, we plot also the electron star solution for comparison. One should be careful in comparing the two, however, as the electron star corresponds to the limit $e\to 0$ and thus cannot be compared directly (i.e., for the same parameter values) to our WKB star. Our convention is to first define the electron star by choosing the \emph{total} charge density $Q$ and the parameter $\hat{m}=m/e\kappa$, where $\kappa$ is the gravitational constant whose value is fixed by the normalization of the action (\ref{sfluid}). For the WKB star, we impose the same value of $Q$, while the value of $m$ is found as $m=\hat{m}e\kappa$ (for WKB star we can control $e$ as an independent parameter). Relative proximity of the solutions for large $N$ seems to confirm that this is a physically meaningful way of comparing the models.

\subsection{Thermodynamics}

We can now use these full solutions to determine the macroscopic characteristics of the dual strongly coupled fermion system. Let us first derive the free energy of the boundary field theory. According to the dictionary, it is equal to the (Euclidean) on-shell action, which contains both bulk and boundary components:
\begin{equation}
\label{frendef}F=S_{bulk}^{on-shell}+S_{bnd}^{on-shell}.
\end{equation}
We have already discussed the bulk action in the previous section. We will again approximate the fermionic contribution (\ref{fermiaction}) by its leading term, the pressure.

In computing the free energy using AdS/CFT a crucial part is often played by boundary terms in the action. It encapsulates the regularizing terms that eliminate $z\rightarrow 0$ divergences, enforces a Dirichlet boundary condition for the gauge field, but it also provides the kernel for the fermionic correlation functions \cite{Leiden:2009,LiuIqbal:2011}
\begin{equation}
\label{sbnd}S_{bnd}=\oint_{\partial AdS}\sqrt{-h}\left(\frac{1}{2}n_\nu F^{\mu\nu}A^\mu+\bar\Psi_+\Psi_-\right),
\end{equation}
with $h$ being the induced metric on the boundary ($h=\frac{1}{z^2}(-1/f(z=0),1,1)$) and $\Psi_+$ and  $\Psi_-$ are radial projections of the wave function as in Eq.~(\ref{proj}). By $\partial AdS$ we have denoted the boundary of the AdS space. Let us now briefly show why these boundary fermion terms do not contribute to the free energy, but that the leading fermion contribution is the (one-loop) effective pressure. Essentially the point is that only normalizable modes of the field are occupied and hence they cannot contribute to the boundary action as they die off too fast. The Dirac field asymptotics at the boundary are given by \cite{csz2010}:
\begin{eqnarray}
\label{psiabnd}\Psi_+=\frac{i\mu\gamma^0}{2m+1}B_-z^{5/2+m}+\ldots, \text{  } & \Psi_-=B_-z^{3/2+m}+\ldots
\end{eqnarray}
At the same time the electromagnetic boundary term reduces to $\Phi\partial_z\Phi \vert_{z=0}=-\mu\rho$, where $\rho$ is the total boundary (not only fermionic) charge density, read off from the subleading ``response'' of the bulk electrostatic potential $\lim_{z\to 0}\Phi(z)=\mu-\rho z+\ldots$. The regularized boundary action now reads
\begin{equation}
\label{sczs}S_{bnd}=\lim_{z_0\to 0}S(z_0)+\lim_{z_0\to 0}\int d^3x\left[\frac{3\mu}{2(2m+1)}\bar{B}_-i\gamma^0B_-z_0^{1+2m}-\frac{1}{2}\mu\rho\right],
\end{equation}
Since $m>-1/2$ is the fermionic unitarity bound in AdS/CFT, the first term always vanishes in the limit $z_0\to 0$. The total on-shell action, i.e. the free energy is therefore
\begin{equation}
\label{sfin}F=\int_{z_0}^{z_H}dzd^3x\sqrt{-g}\left[R+6+\frac{ze^{h/2}qn\Phi}{2\sqrt{f}}+p\right]-\frac{1}{2}\mu\rho
\end{equation}

\subsection{Constructing the phase diagram: Quantum corrections imply a first order thermal phase transition to AdS-RN}
\label{FirstOrderPT}

The condensed matter context in which we are discussing AdS/CFT is that of an emergent finite density fermionic ground state out of an UV CFT. In the deep UV or at very high temperatures $T/\mu$ the chemical potential should be negligible and we should recover as the preferred groundstate the UV CFT at finite $T/\mu$. The gravitational dual of this is the AdS-RN black hole. It describes a conformal  critical phase with no Fermi surfaces. As we lower $T/\mu$ an instability should set in towards a state with a finite occupation number of fermions. In the probe analysis one indeed finds several normalizable wavefunctions signalling the existence of states with distinct occupation numbers. They are the bulk counterpart of the existence of non-Fermi-liquid Fermi surfaces \cite{Vegh:2009,Leiden:2009,Faulkner:2009,Liu:2011}. A crucial qualitative aspect is that due to their fermionic nature the wavefunctions of these normalizable modes can {\em never} ``grow''. From a microscopic point of view it therefore appears that any fermion driven phase transition cannot be second order. In the fluid limit, however, the transition was found to be third order. There is no conflict because new analytic behavior can emerge in the fluid scaling limit where the {\em number of Fermi surfaces} is taken to infinity.\footnote{Note that there is a crucial subtlety in the fluid limit in AdS/CFT with a flat Minkowski-space boundary. Normally one needs a thermodynamic ``fluid'' limit to even be able to discuss the notion of a phase transition. In global AdS, or conventional Tolman-Oppenheimer-Volkov neutron stars, a bound on the number of radial modes, implies a countable number of states. However, this is not so in AdS/CFT with a flat Minkowski-space boundary. For each radial mode there is still a formal infinite number of modes distinguished by the transverse momentum. The phase transition discussed here is where one considers $N/V_{transverse}\rightarrow\infty$. It restores one's intuition that the emergence of {\em each single Fermi surface} dual to each single radial mode is associated with a macroscopic phase transition. We thank Sean Hartnoll for emphasizing this.} It does mean that one has to be quite careful in the fluid limit as for fermions these corrections can change macroscopic quantities. For any finite number of Fermi surfaces we should discover a first order transition. We did indeed find this earlier in the Dirac hair approximation valid for $N_{WKB}=1$ \cite{csz2010}. With the WKB construction put forward here, we will show that this is indeed so for any finite $N_{WKB}$.

\begin{figure}[tb]
\begin{center}
(A)\includegraphics[width=0.46\linewidth]{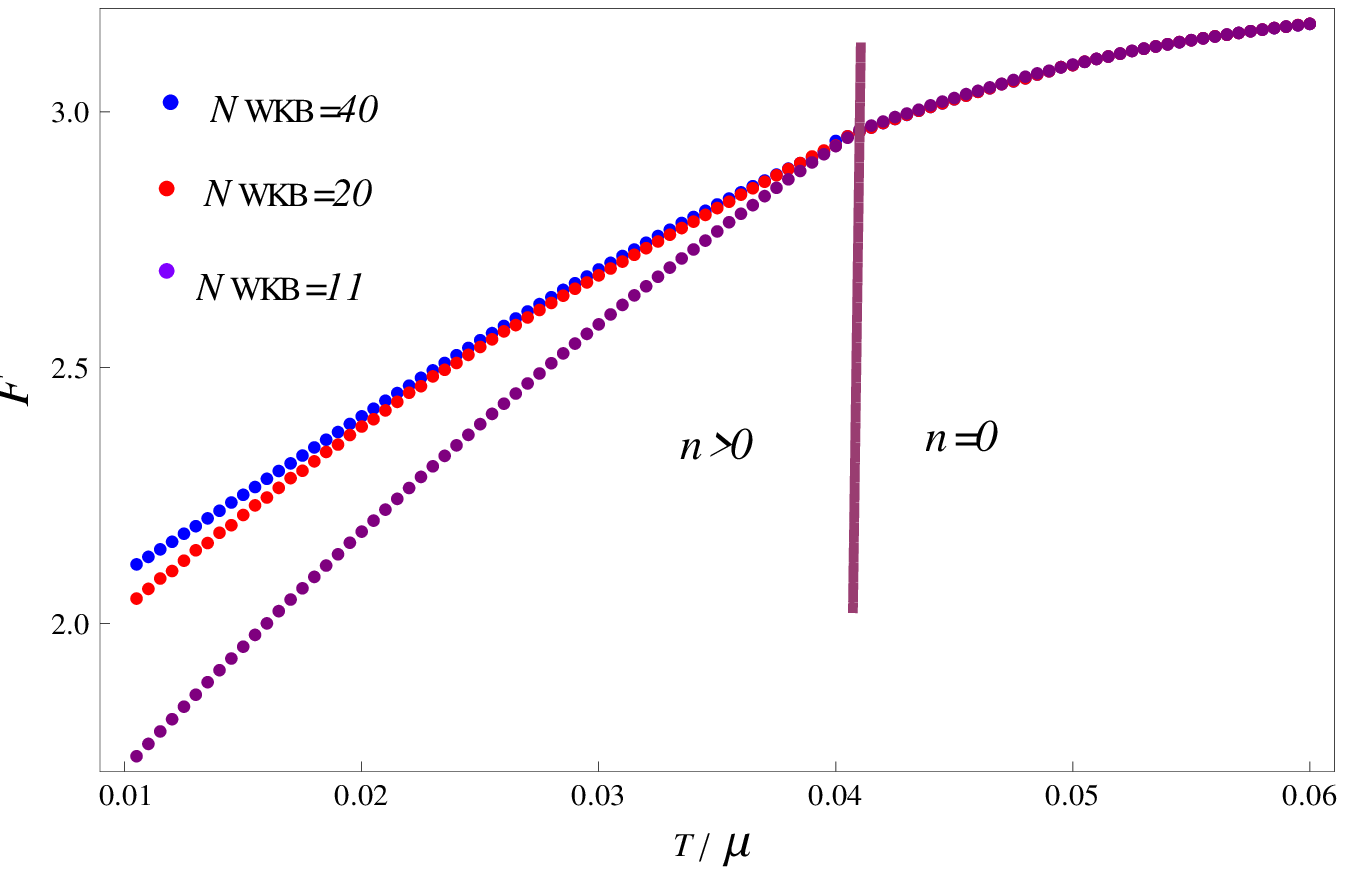}
(B)\includegraphics[width=0.46\linewidth]{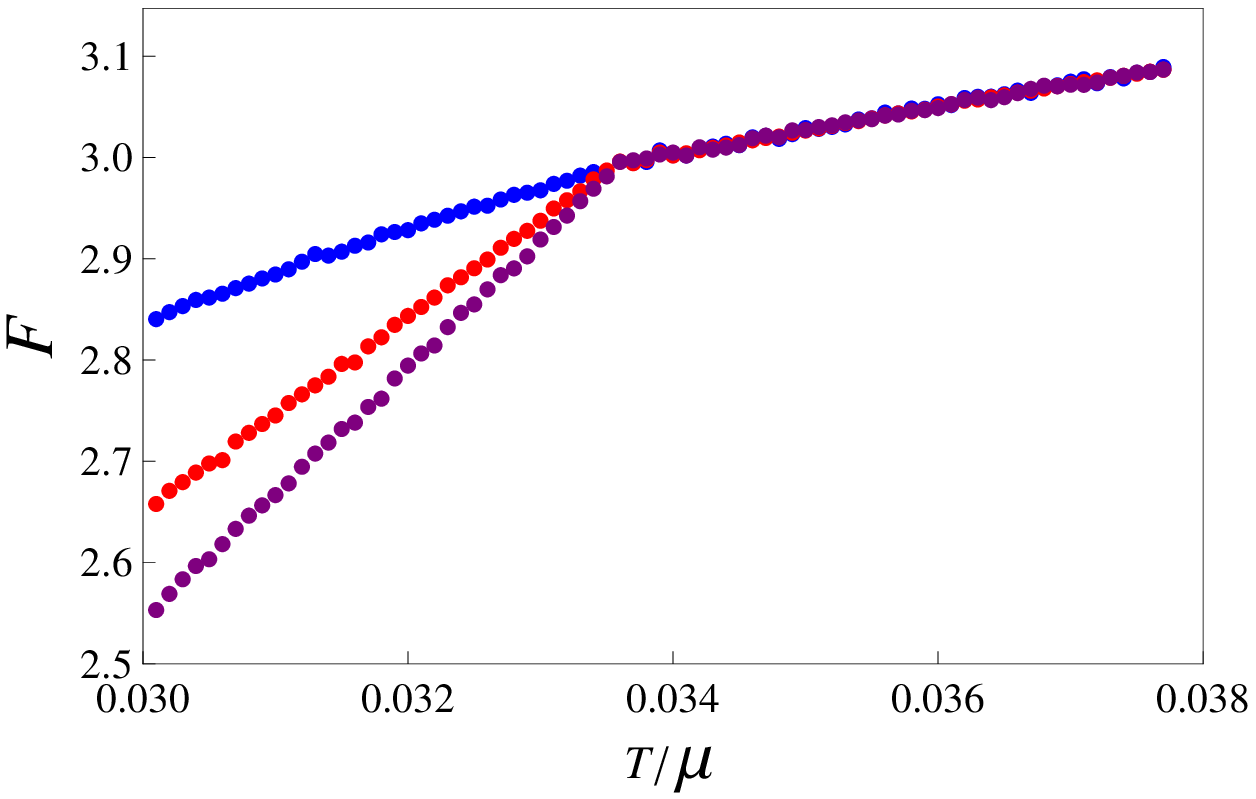}
\caption{Free energy as a function of temperature $F(T)$. The abrupt change of the derivative signifies the first order transition between the finite density phase and the pure black hole (with zero bulk fermion density), in line with the analytical prediction of the first order transition from the second term in the bulk free energy in Sec.~\ref{FirstOrderPT}. We show the calculations for three different values $(\mu,e,m)$ of the system parameters: $(1.7,3,0.1)$, $N_{WKB}(T=0)=40$ in blue, $(1.7,10,0.1)$, $N_{WKB}(T=0)=20$ in red and $(1.7,10,0.7)$, $N_{WKB}(T=0)=11$ in violet. Notice how the slope of $F$ in the low-temperature phase decreases as the number of levels increases: for $N_{WKB}\to\infty$ we reach the electron star limit when the transition becomes continuous. Panel (B) shows the vicinity of the critical temperature for three sets of parameter values, to make the cusp in $F(T)$ clearly visible. In the high temperature (RN) phase the curves $F(T)$ fall on top of each other as one expects for the RN black hole with $n=0$. The behavior in the low-temperature phase (with non-zero density) is different for the three curves as the value of the charge affects the behavior of the bulk fermions. For presentation purposes, the curves have been rescaled to the same transition temperature; in general, however, $(T/\mu)_c$ is \emph{not} universal and will differ for different corners of the parameter space.}
\label{figfren1}
\end{center}
\end{figure}

Fig.~\ref{figfren1} shows the behavior of the free energy $F(T/\mu)$ of the WKB corrected star construction for different parameters $e, m$, corresponding to a different number of levels $N_{WKB}$ (which roughly equals the number of Fermi surfaces). In the high temperature phase the preferred state with lowest $F(T/\mu)$ is that of the pure AdS-RN. Since there are no occupied fermionic states it is independent of the fermion charge and mass. In the low temperature phase the preferred phase is the WKB star. Where the phase transition occurs, one immediately sees the characteristic first order cusp in $F(T/\mu)$ whose non-analyticity indeed becomes clearer as $N_{WKB}$ decreases. The panel (B) of the figure makes this clear by showing the vicinity of the phase transition.

The first order nature of the phase transition can in fact be understood analytically with this WKB construction. The argument is along similar lines as for the fluid limit of the electron star \cite{Hartnoll:phtr}. Assuming that the transition is dominated by the behavior of the fermions and that the contribution of the geometry change due to backreaction is small near the critical temperature, the relevant part of the free energy of the system is given by
\begin{equation}
\label{frenpress}F_{Fermi}\approx\int_0^{z_H}p=\frac{e\mu}{2}\int_{z_*}^{z_{**}}(1-z)n+\frac{e\mu}{2}\int_{z_*}^{z_{**}}\frac{\tilde{M}^2e^{-h}}{z\sqrt{\tilde{M}^2+e^h\tilde{E}^2}}\equiv F_{Fermi}^{fluid}+\Delta F_{Fermi}
\end{equation}
Starting from low temperatures and nonzero $n$, at the transition point the bulk density $n$ vanishes. In the WKB construction that means that the turning points coincide: $z_*\to z_{**}$. The first, ''fluid limit'' term $F_{Fermi}^{fluid}$ in (\ref{frenpress}) is proportional to $\Phi n$ and it is analyzed in detail in \cite{Hartnoll:phtr}. It yields the scaling ${F}_{Fermi}^{fluid}\sim (T-T_c)^3$. This indicates a third order transition at the semi-classical level. The new, second, quantum term will change this, however. The vanishing of the classically allowed region means $\tilde{E}\approx\tilde{M}$ in the whole (narrow) region $z_*<z<z_{**}$. One can thus expand $\tilde{E}=\tilde{M}+\delta z\times\delta\tilde{E}/\delta z+\ldots$ and analyze the leading terms in $\delta z$. It is easy to see that its expansion starts from a constant. Since for vanishing $\delta z$ the density can be assumed constant throughout the WKB star, we estimate the integral in $\Delta F_{Fermi}$ as
\begin{equation}
\Delta F_{Fermi}\approx\frac{\Phi\tilde{M}^2e^{-h}}{\sqrt{\tilde{M}^2+e^h\tilde{E}^2}}\delta z=\left(\mathrm{const.}+O\left(\delta z\right)\right)\delta z,
\end{equation}
where $\delta z=z_{**}-z_*$. Therefore, the second term scales as $\Delta F_{Fermi}\sim\delta z$. Now, for a vanishing bulk charged fluid/emerging charged black hole, the principle of detailed balance predicts that the charge of the former equals the charge of the latter: $n\delta z=n_{BH}\delta z_H$, where the charge densities of the bulk and the black hole are $n$ and $n_{BH}$, respectively, and $\delta z_H$ is the change in the position of the black hole horizon. Since the densities can be assumed constant for vanishing $\delta z$ and $\delta z_H$, we find $\delta z\sim\delta z_H\sim T-T_c$. We can now write $F_{Fermi}=F_{Fermi}^{fluid}+\Delta F_{Fermi}$. We know that $F_{Fermi}^{fluid}\sim (T-T_c)^3$ \cite{Hartnoll:phtr}, but we have now shown that
\begin{equation}
\label{first}\Delta F_{Fermi}\sim T-T_c.
\end{equation}
At the quantum level the \emph{transition is always of first order}. The quantum correction is subleading at general $T$ values, but becomes leading as the phase transition point is approached. Finally, we remark that, if one considers the bulk free (or internal) energy $\int dz\mathcal{E}$ given in Eq.~(\ref{energy}) using the similar scaling reasoning, one arrives at the same conclusion: $F\sim T-T_c$. This confirms the intuition that the bulk and boundary thermodynamics are equivalent at leading order, i.e. the difference $F_{bulk}-F$ does not contain first-order terms in $T-T_c$ and thus does not change the order of the transition. Now the exact free energy differs from our WKB star calculation, as we have assumed that the correction to the fluid limit is fully captured by the correction to pressure. However, an additional term in $F$ \emph{cannot decrease the order of the transition}: it can introduce new singularities (of some order $\alpha$, scaling as $(T-T_c)^\alpha$) but cannot cancel out the term.

The numerics just confirms this analytic prediction of a first order phase transition. The field theory interpretation of the discontinuous nature of the transition to a phase with Fermi surfaces is simple: fermions do not break any symmetry but the discharge of the black hole does signify that the ground state is reconstructed due to the formation of a rigid Fermi surface. The only way to reconstruct the ground state without breaking any symmetries is precisely the first order transition of the density van der Waals liquid-gas type. This is the macroscopic counterpart to the probe analysis where the Grassman nature of fermions Pauli blocks the growing of mode functions. A van der Waals liquid-gas first order type transition is indeed seen in \cite{csz2010} for the first order transition from $N_{WKB}=1$ Dirac hair state to AdS-RN. The confusing point was that electron star/AdS-RN transition valid in the strict $N_{WKB}\rightarrow \infty$ fluid limit was found to be third order \cite{Hartnoll:phtr,Larus}. Here we show that this change in the nature of the phase transition is an artifact of this $N_{WKB}\rightarrow \infty$ limit. Instead the expected first order behavior is recovered for any finite value of $N_{WKB}$.

\section{Discussion and conclusions}

\begin{figure}[ht!]
\begin{center}
\includegraphics[width=10cm]{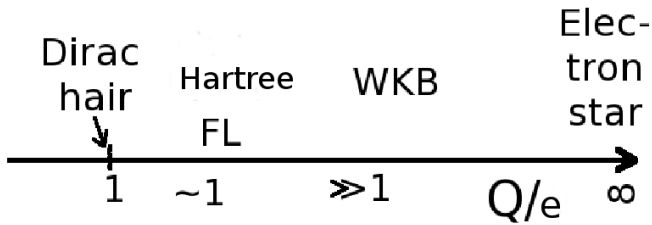}
\caption{Applicability of various approximations as a function of the ratio of the fermion charge and the total charge of the system, $Q/e$: Dirac hair, confined Fermi liquid, our present WKB-model, electron star. Dirac hair and electron star are the simplest and most flexible approximations but limited to the extreme ends of the $Q/e$ axis. Compare also to Fig.~10 in \cite{Leiden:2011}.}
\label{figapp}
\end{center}
\end{figure}

In this paper we have constructed the WKB star as an improved semiclassical model of holographic fermions in AdS${}_4$ space, aimed at understanding the phase diagram of strongly coupled Fermi and non-Fermi liquids. The model combines a WKB approximation with a Hartree summation to approximate a finite $N_{WKB}$ charged fermion state in AdS coupled to both gravity and electromagnetism. The dominant effect is a quantum correction to the pressure and energy density ("vacuum polarization") of the conventional $N_{WKB}\rightarrow \infty$ classical model -- the electron star. This finite $N_{WKB}$ approach has allowed us to address the intermediate fermion charges which cannot be modeled satisfyingly with any of the previously used models.

By studying the free energy of the system we can now construct the full phase diagram of the system. Most importantly, we find a \emph{universal first order phase transition from a finite density to a zero density (Reissner-Nordstr\"{o}m, quantum critical) phase}. The  discontinuity of the density comes from the quantum term in the internal energy. This term is always present but its relative contribution to the free energy decreases with the inverse of the number of radial modes $N_{WKB}$. The extreme limit $N_{WKB}\to\infty$  reproduces the unexpected third order continuous phase transition found in \cite{Larus,Hartnoll:phtr}. Nevertheless, in any real system with finite fermion charge the discontinuity will be present, which fits into the general expectation that the thermal phase transition of a fermionic system should be of the van der Waals (liquid-gas, Ising) type.

So far three distinct approaches aiming at capturing the stable phases of holographic fermionic matter have appeared: the electron star \cite{Hartnoll:es}, Dirac hair \cite{csz2010} and the confined Fermi liquid model \cite{Sachdev}. The electron star is essentially a charged fermion rewriting of the well-known Oppenheimer-Volkov equations for a neutron star in AdS background. The bulk is thus modeled as a semiclassical fluid. It is a controlled approximation in the certain limit of the parameter values. The mystery is its field theory dual: it is a hierarchically ordered (infinite) multiplet of fermionic liquids with stable quasiparticles \cite{Hartnoll:2011}. On the other end of the spectrum is Dirac hair, which reduces the bulk fermion matter to a single radial harmonic. The Dirac hair approach is based on the truncation of the full non-local equations of motion. As a consequence the field theory dual is a single Fermi liquid, however its gravitational consistency properties are not yet fully understood. In \cite{Leiden:2011} we have shown that Dirac hair and electron star can be regarded as the extreme points of a continuum of models, dialing from deep quantum -– a single radial mode -— to a classical regime —- a very large occupation number -— in the bulk. They correspond to two extreme "phases" in the field theory phase diagram: a multiplet of a very large number of Fermi liquids and a single Fermi liquid. The third approach \cite{Sachdev} performs a Hartree summation of the exact quantum mechanical wave functions to capture the fermion density. While the paper \cite{Sachdev} applies the Hartree method to a specific model (confined Fermi liquid, where the confinement is intrduced through modifying the bulk geometry), the main idea can be used in any background. This approach is more general then the single-particle approach of \cite{csz2010} and it naturally extends the single harmonic Dirac hair state with a single Fermi surface to a state with multiple Fermi surfaces. Our main motivation is to construct a complementary model that extends from the other end --- the semi-classical fluid --- down to a state with a countable but large number of Fermi surfaces. We aim for a system which is general enough to encompass the middle ground between extreme quantum and extreme classical regimes in the original deconfined setup. In the recent model of "quantum electron star" \cite{McGreevy:2012} the same goal is set but the method used is different and is based on the deconfined limit of \cite{Sachdev}.

In constructing the WKB star, we were also guided by the strengths and weaknesses of these existing models. On the one hand, the Dirac hair is a fully quantum-mechanical model which shows its strength in particular near the boundary (the ultraviolet of the field theory) but becomes worse in the interior, i.e. close to the horizon (the infrared of the field theory) where density is high and the resulting state of matter cannot be well described by a single wave function. On the other hand, the electron star yields a very robust description of high-density matter in the interior but its sharp boundary at some radius $r_c$ is clearly incompatible with a fully quantum description. It is thus obvious that the physically interesting model lies somewhere in-between the two approaches.

How to relate the electron star \cite{Hartnoll:es}, Dirac hair \cite{csz2010} and the (confined) Hartree Fermi liquid \cite{Sachdev} to our new phase diagram? All models use the same microscopic action for a Dirac fermion with charge $e$ and mass $m$, but the system is approximated in different ways. The electron star is the fluid limit of the equations of motion, yielding the Openheimer-Volkov equations in the bulk. As explained in \cite{Leiden:2011}, this approximation is valid in the limit of infinite occupation number $N_{WKB}\rightarrow\infty$, $e \rightarrow 0$ with the total charge density fixed $Q=N_{WKB}e$. In addition, the mass $m\rightarrow 0$ while $\hat{m}=m/\sqrt{N_{WKB}}e$ is fixed. The Dirac hair departs from the opposite limit, treating the bulk fermion as a single collective excitation with $N_{WKB}=1$. To obtain a macroscopic charge density one essentially has to take $e \gg 1$. Finally, the confined Fermi liquid of \cite{Sachdev} and its deconfined version \cite{McGreevy:2012} improve on the Dirac Hair by using a standard Hartree summation of the non-interacting bulk Fermi gas. It works for all $N_{WKB}\sim{\cal O}(1)$ and this significantly increases the region of applicability but at the cost of substantial practical complications, in particular if one wishes to take into account the backreaction on the metric \cite{McGreevy:2012}. Our model takes a similar summation approach but simplifies the wave function calculation drastically by using the WKB approximation. This inherently assumes semiclassical dynamics and large number of energy levels $N_{WKB}\gg 1$ in the bulk. The WKB star is thus independent of \cite{csz2010} but draws heavily on the electron star and the dialing concept of \cite{Leiden:2011}. Since we do \emph{not} make the assumption of zero energy spacing $N_{WKB}\rightarrow\infty$ necessary for the fluid approximation, our model thus works well in the intermediate regime where $N_{WKB}$ is finite but large compared to unity. This message is illustrated in Fig. \ref{figapp},emphasizing the singular nature of both the electron star and the Dirac hair.

One obvious downside of the WKB star is that the WKB approximation breaks down when $N_{WKB}$, the occupation number, is low. In particular, it means that the accuracy of our method is lowest close to the phase transition to the RN phase. However, for reasons outlined in the Section VB, we can argue that the order of the phase transition cannot change, i.e. the first-order singularity in the free energy will not be canceled out by the corrections to WKB. Our treatment is an improvement over the strict $N_{WKB}\to\infty$ limit of the electron star model used in \cite{Hartnoll:phtr,Larus} to analyze the phase transition, however it remains a task for further work to approach the transition point with a more accurate method which is not limited to large occupation numbers. The recent paper \cite{newref} constructs a solution with finite fermion density in AdS${}_4$ without using WKB: this turns out to be much more involved, but allows one to move away from the large $N_{WKB}$ regime.

The natural next step departing from this WKB treatment is to employ a fully quantum-mechanical density functional method. It is, in fact, not a significant complication compared to the approach of this paper: the recipe for computing the density $n$ will be replaced by a somewhat more complicated functional of the gauge field and the metric, which needs to be computed iteratively. We anticipate that this will not alter the qualitative picture, although the number might change significantly. The main conclusion of our paper is that the singular fluid limit of bulk fermions when coupled to AdS gravity can lead to macroscopically anomalous results. Finite $N_{WKB}$ corrections are crucial to get the correct answer.

\section*{Acknowledgments}

\noindent This research was supported in part by a Spinoza Award (J.~Zaanen) from the Netherlands Organization for Scientific Research (NWO) and the Dutch Foundation for Fundamental Research on Matter (FOM). We are grateful to H.~Liu, S.~S.~Gubser and A.~Karch for valuable discussions.


\begin{thebibliography}{99}

\bibitem{Hartnoll:2010}
S.~A.~Hartnoll,
"Lectures on holographic methods for condensed matter physicists",
Class.~Quant.~Grav. {\bf 26}, 224002 (2009)
[arXiv:0903.3246[hep-th]].

\bibitem{McGreevy:2010}
J.~McGreevy,
"Holographic duality with a view toward many-body physics", Adv. in High En. Phys., {\bf 2010}, Article ID 723105,  (2010)
arXiv:0909.0518[hep-th].

\bibitem{heavy}
H. v. L\"{o}hneysen, A. Rosch, M. Vojta, P. W\"{o}lfle
"Fermi-liquid instabilities at magnetic quantum phase transitions"
Rev. Mod. Phys. {\bf 79}, 1015 (2007)
[arXiv:cond-mat/0606317 [cond-mat]].

\bibitem{strange}
M.~Gurvitch, A.~T.~Fiory,
"Resistivity of $La1.825$$Sr0.175$$CuO4$ and $YBa2$$Cu3$$O7$ to 1100 K: Absence of saturation and its implications",
Phys.\ Rev.\ Lett.\  {\bf 59}, 1337 (1987).

\bibitem{Gauntlett:1}
A.~Donos, J.~P.~Gauntlett, N.~Kim, O.~Varela,
"Wrapped M5-branes, consistent truncations and AdS/CMT",
JHEP {\bf 1012}, 003 (2010)
[arXiv:1009.3805[hep-th]]

\bibitem{Gubser}
O.~DeWolfe, S~.S.~Gubser, C.~Rosen,
"Fermi surfaces in maximal gauged supergravity",
Phys.~Rev.~Lett. {\bf 108}, 251601 (2012)
[arXiv:1112.3036[hep-th]].

\bibitem{Gauntlett:2}
A.~Donos, J.~P.~Gauntlett, C.~Pantelidou,
"Magnetic and electric AdS solutions in string- and M-theory",
[arXiv:1112.4195[hep-th]]

\bibitem{Vegh:2009}
H.~Liu, J.~McGreevy, D.~Vegh,
"Non-Fermi liquids from holography",
Phys.\ Rev.\ D {\bf 83}, 065029 (2011)
[arXiv:0903.2477[hep-th]].

\bibitem{Leiden:2009}
M.~\v{C}ubrovi\'{c}, J.~Zaanen and K.~Schalm,
"String Theory, Quantum Phase Transitions and the Emergent Fermi-Liquid",
Science {\bf 325}, 439 (2009),
[arXiv:0904.1993[hep-th]].

\bibitem{Hartnoll:es} S.~A.~Hartnoll, A.~Tavanfar,
"Electron stars for holographic metallic criticallity",
Phys.\ Rev.\ D {\bf 83}, 046003 (2011)
[arXiv:1008.2828[hep-th]].

\bibitem{GauntlettSonner}
J.~P.~Gauntlett, J.~Sonner, D,~Waldram, "Universal fermionic spectral functions from string theory",
Phys. Rev. Lett. {\bf 107}, 241601 (2011)
[arXiv:1106.4694 [hep-th]].

\bibitem{Hartnoll:phtr} S.~A.~Hartnoll, P.~Petrov,
"Electron star ar birth: A Continuous Phase Transition at Nonzero Density",
Phys.\ Rev.\ Lett. {\bf 106}, 121601 (2011)
[arXiv:1011:6469[hep-th]]

\bibitem{csz2010}
M.~\v{C}ubrovi\'{c}, J.~Zaanen, K.~Schalm,
"Constructing the AdS dual of a Fermi liquid: AdS Black holes with Dirac hair",
JHEP {\bf 2011}, 17 (2011).
[arXiv:1012.5681[hep-th]]

\bibitem{Sachdev}
S.~Sachdev,
``A model of a Fermi liquid using gauge-gravity duality,''
Phys.\ Rev.\ D {\bf 84}, 066009 (2011).
[arXiv:1107.5321 [hep-th]]

\bibitem{ActionFluid}
R.~C.~Tolman,
``Static solutions of Einstein's field equations for spheres of fluid,''
Phys.\ Rev.\ D {\bf 55}, 364 (1939).

\bibitem{McGreevy:2012}
A.~Allais, J.~McGreevy and X.~Josephine Suh,
"Quantum electron star,"
Phys.~Rev.~Lett {\bf 108}, 231602
[arXiv:1202.5308[hep-th]]

\bibitem{Hartnoll:2011}
S.~A.~Hartnoll, D.~M.~Hofman and D.~Vegh,
``Stellar spectroscopy: Fermions and holographic Lifshitz criticality,''
JHEP {\bf 1108}, 96 (2011).
[arXiv:1105.3197 [hep-th]]

\bibitem{Leiden:2011}
M.~\v{C}ubrovi\'{c}, Y.~Liu, K.`Schalm, Y.-W.~Sun and J.~Zaanen,
``Spectral probes of the holographic Fermi groundstate: Dialing between the electron star and AdS Dirac hair''
Phys.\ Rev.\  {\bf D84}, 086002 (2011),
[arXiv:1106.1798 [hep-th]]

\bibitem{Liu:2011}
N.~Iqbal, H.~Liu and M.~Mezei,
"Semi-local quantum liquids",
[arXiv:1105.4621[hep-th]]

\bibitem{Review:2010}
S.~A.~Hartnoll, J.~Polchinski, E.~Silverstein, D.~Tong,
"Towards strange metallic holography",
JHEP {\bf 1004}, 120 (2010).
[arXiv:0912.1061[hep-th]]

\bibitem{Faulkner:2009}
T.~Faulkner, H.~Liu, J.~McGreevy and D.~Vegh,
"Emergent quantum criticality, Fermi surfaces, and AdS2", Phys. Rev. D {\bf 83}, 125002 (2011)
[arXiv:0907.2694 [hep-th]].

\bibitem{Leiden:2010}
E.~Gubankova, J.~Brill, M.~\v{C}ubrovi\'c, K.~Schalm, P.~Schijven, J.~Zaanen,
"Holographic fermions in external magnetic fields", Phys.\ Rev. D {\bf 84}, 106003 (2011)
[arXiv:1011.4051[hep-th]].

\bibitem{epsilon}
D.~F.~Mross, J.~McGreevy, H.~Liu, T.~Senthil,
"A controlled expansion for certain non-Fermi liquid metals",
Phys.\ Rev.\ B {\bf 82}, 045121 (2010)
[arXiv:1003.0894[hep-th]].

\bibitem{Horowitz}
G.~Horowitz, A.~Lawrence, E.~Silverstein,
"Insightful D-branes",
JHEP {\bf 0907}, 057 (2009)
[.arXiv:0904.3922[hep-th]]

\bibitem{ll3}
L.~D.~Landau and E.~M.~Lifshitz,
"Quantum Mechanics. Nonrelativistic theory",
Nauka, Moskva, 1989.

\bibitem{Faulkner:2010}
T.~Faulkner, N.~Iqbal, H.~Liu, J.~McGreevy and D.~Vegh
"Strange metal transport realized by gauge/gravity duality",
Science {\bf 329}, 1043 (2010)
[arXiv:1003.1728[hep-th]].

\bibitem{Hartman:2010}
T.~Hartman, S.~A.~Hartnoll,
"Cooper pairing near charged black holes",
[arXiv:1003.1918[hep-th]].

\bibitem{Iqbal:2009}
N.~Iqbal and H.~Liu,
"Real-time response in AdS/CFT with application to spinors",
Fortsch.\ Phys.  {\bf 57}, 367 (2009).
[arXiv:0903.2596 [hep-th]]

\bibitem{Landau9}
L.~D.~Landau and E.~M.~Lifshitz.,
"Statistical Physics 2",
Nauka, Moskva, 1978.

\bibitem{Iqbal:2010}
N.~Iqbal, H.~Liu, M.~Mezei, Q.~Si, "Quantum phase transitions in
holographic models of magnetism and superconductors", Phys.\ Rev.\
D {\bf 82}, 045002 (2010)
[arXiv:1003.0010[hep-th]].

\bibitem{voskresenie}
D.~N.~Voskresenskii and A.~V.~Senatorov,
"Vacuum reconstruction in strong electric and gravitational fields",
Sov.~J.~Nucl.~Phys {\bf 36}, 208 (1982).

\bibitem{SZ}
J.~H.~She, J.~Zaanen,
"BCS Superconductivity in Quantum Critical Metals",
Phys. Rev. {\bf B80}, 184518 (2009).

\bibitem{Larus}
V.~Giangreco~M.Puletti, S.~Nowling, L.~Thorlacius and T.~Zingg,
"Holographic metals at finite temperature",
JHEP {\bf 2011}, 117 (2011)
[arXiv:1011.6261[hep-th]]

\bibitem{polchin}
T.~Faulkner and J.~Polchinski,
"Semi-holographic Fermi liquids",
JHEP {\bf1106}, 012 (2011)
[arXiv:1001.5049[hep-th]]

\bibitem{Sachdev:2}
L.~Huijse and S.~Sachdev,
"Fermi surfaces and gauge-gravity duality",
Phys.~Rev.~D {\bf 84}, 026001 (2011).
[arXiv:1104.5022[hep-th]]

\bibitem{LiuIqbal:2011}
N.~Iqbal, H.~Liu, 2011.
"Lectures on holographic non-Fermi liquids and quantum phase transitions",
[arXiv:1110.3814[hep-th]]

\bibitem{thesis}
M.~\v{C}ubrovi\'{c},
\emph{Holography, Fermi surfaces and criticality},
Casimir PhD series, Leiden Univeristy, 2013.

\bibitem{newref}
A.~Allais,~J.~McGreevy, 2013.
"How to construct a gravitating quantum electron star",
[arXiv:1306.6075[hep-th]]

\end{thebibliography}
\end{document}